\documentclass[prl,aps,twocolumn,superscriptaddress]{revtex4}
\usepackage{amsfonts}
\usepackage{amsmath}
\usepackage{amssymb}
\usepackage{graphicx}
\usepackage{txfonts}
\usepackage{bm}

\begin{document}

\title{Symmetry-Projected Weakly Compatible Multiparameter Quantum Sensing}
\author{G. R. Jin}
\email{grjin@zstu.edu.cn}
\affiliation{Zhejiang Key Laboratory of Quantum State Control and Optical Field
Manipulation, Department of Physics, Zhejiang Sci-Tech University, 310018
Hangzhou, China}
\author{Z. Y. Zhou}
\email{zheng-yang.zhou@zstu.edu.cn}
\affiliation{Zhejiang Key Laboratory of Quantum State Control and Optical Field
Manipulation, Department of Physics, Zhejiang Sci-Tech University, 310018
Hangzhou, China}
\author{W. Yang}
\email{wenyang@csrc.ac.cn}
\affiliation{Beijing Computational Science Research Center, Beijing 100193, China}
\date{\today }

\begin{abstract}
Achieving joint quantum-enhanced precision in multiparameter sensing
requires both high sensitivity and measurement compatibility. These two
aspects are characterized by the quantum Fisher information matrix (QFIM)
and the Uhlmann curvature matrix (UCM), respectively, with weak
compatibility corresponding to the vanishing of the relevant UCM elements.
Here, we develop a symmetry-projection framework that classifies phase
generators into subspace-preserving and subspace-changing sectors. For probe
states confined to a symmetry subspace, symmetry projection imposes a common
block-diagonal structure on the QFIM and UCM, rendering cross-sector
parameters simultaneously free from information cross-talk and measurement
incompatibility. When the subspace-changing generators act as scalars within
the occupied subspace, the corresponding QFIM block reduces to four times
the symmetrized covariance matrix, even for mixed probe states. For
parity-protected collective $\mathrm{SU}(2)$ systems, this structure singles
out the transverse anti-squeezed quadrature and the longitudinal mean-spin
direction as natural optimal sensing axes. Applied to a dissipative one-axis
twisting model, the dynamically generated probe state exhibits identically
vanishing UCM elements for transverse--longitudinal parameter pairs, while
maintaining nearly balanced, Heisenberg-scaled QFIM components over a broad
transient window. Our work opens a route to symmetry-protected, weakly
compatible multiparameter sensing in interacting quantum many-body systems.
\end{abstract}

\maketitle

%\pacs{03.67.-a, 03.67.Mn, 06.20.Dk, 42.50.St}

%======================================================================%

Multiparameter quantum metrology targets the simultaneous estimation of multiple physical parameters~\cite{Holevo1982,Hayashi2005}, a capability pivotal to vector-field sensing, multidimensional imaging, multiphase estimation, and quantum process characterization~\cite{Humphreys2013,Baumgratz2016,Szczykulska2016,Albarelli2020,Kau23}. Compared with single-parameter metrology~\cite{helstrom1976quantum,Yurke1986,braunstein1994statistical,giovannetti2004quantum,giovannetti2011advances,Kitagawa,Wineland1,Wineland1994,Wineland2,Wineland3,ma2011quantum,pezze2018quantum,PezzePRL09}, it faces two central challenges: information cross-talk encoded in off-diagonal elements of the quantum Fisher information matrix (QFIM) and measurement incompatibility~\cite{Matsumoto2002,Pezze2017,Sidhu21,Dem20,Ragy2016,Yamagata,Yang2019,AlbarelliJPA2019,Gill2000,Liu2020,SidhuKok2020,XMLu2021}. The latter is characterized by the mean Uhlmann curvature matrix (UCM), with elements
$\mathcal{I}_{\mu\nu} = \frac{1}{2i}\mathrm{Tr}(\hat{\rho}[\hat{L}_\mu,\hat{L}_\nu])$, defined via the probe state $\hat{\rho}$ and symmetric logarithmic derivatives (SLDs) $\hat{L}_{\mu}$ associated with the parameters $\theta_{\mu}$. Weak compatibility requires the relevant UCM elements to vanish, $\mathcal{I}_{\mu\nu}=0$, in which case the Holevo bound coincides with the SLD quantum Cram\'{e}r--Rao bound (QCRB), rendering the latter asymptotically attainable~\cite{Matsumoto2002,Pezze2017,Sidhu21,Dem20,Ragy2016,Yamagata,Yang2019,AlbarelliJPA2019,Gill2000,Liu2020,SidhuKok2020,XMLu2021}. Achieving joint quantum-enhanced precision therefore requires not only high quantum Fisher information (QFI)~\cite{Hyllus,Reilly}, but also structural mechanisms that suppress information cross-talk while enforcing weak compatibility.

Several strategies have been proposed to address these difficulties, such as
engineering commuting generators~\cite{Reilly}, embedding parameters into
enlarged operator manifolds~\cite{Cao2025}, or utilizing mode entanglement
across distributed sensor arrays~\cite{Fadel22,LiScience26,Gessner2020}.
When such compatible encodings cannot be realized, measurement
incompatibility generally prevents the individual SLD-QCRB from being attained
simultaneously and imposes precision trade-offs~\cite{Matsumoto2002,Pezze2017,Sidhu21,Dem20,Ragy2016,Yamagata,Yang2019,AlbarelliJPA2019,Gill2000,Liu2020,SidhuKok2020,XMLu2021}%
. While optimized collective measurements and Holevo-type bounds
characterize the ultimate precision limits in such incompatible regimes~\cite%
{Pezze2017,WangPRL2026,YanPRA2019}, many protocols still rely on ancillary
degrees of freedom~\cite{Liu17,Yang2022SU2,HouPRL21,Valahu2025,Hou2021},
spatial partitioning, or tailored system optimization~\cite%
{Cao2025,Fadel22,LiScience26}. Consequently, it remains an open question
whether the intrinsic symmetry of a self-contained collective probe can
naturally enforce both information-sector decoupling and weak compatibility,
while reducing the relevant QFIM to directly accessible many-body
observables.

This question is motivated by two considerations. First, symmetry-enhanced
optimal measurements in single-parameter estimation~\cite%
{Xing2024,Frerot,Nolan} and symmetry-based analyses of multiparameter
estimation~\cite{Miyazaki2022,Wang2024antiunitary,Goldberg2021} highlight
symmetry as a valuable resource for quantum metrology. Second, collective $%
\mathrm{SU(2)}$ symmetry underlies a broad class of two-mode interferometers
and collective-spin sensors~\cite%
{Yurke1986,Wineland1,Wineland1994,pezze2018quantum}. The $\mathrm{SU(2)}$
algebra, however, imposes a stringent constraint on multiparameter encoding,
since any two nonparallel collective-spin generators fail to commute. This
raises a fundamental question: Without introducing additional degrees of
freedom or spatial partitioning, can a single symmetry-projected probe state
nevertheless render such noncommuting encodings weakly compatible?

\begin{figure}[hptb]
\centering
\includegraphics[width=\columnwidth]{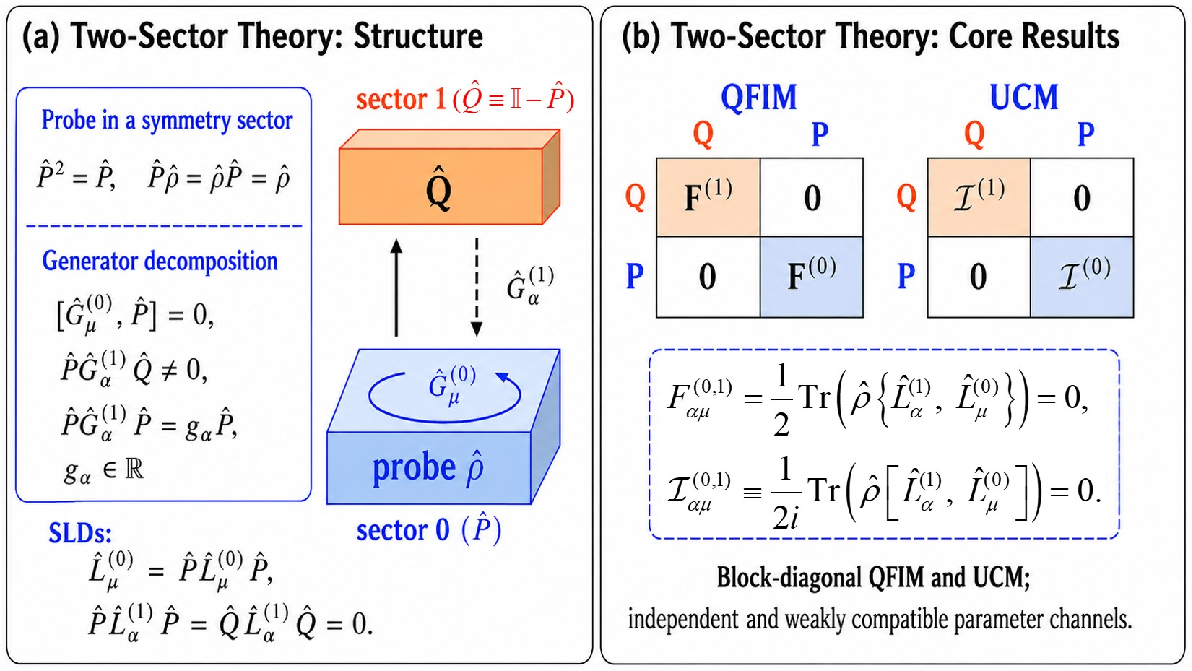}
\caption{Two-sector symmetry-projection mechanism. (a) Decomposition into
the probe-supporting sector~$0$, associated with the projector $\hat P$, and
its orthogonal complement, sector~$1$, with $\hat Q\equiv\mathbb{I}-\hat P$.
The probe satisfies $\hat P\hat\protect\rho=\hat\protect\rho\hat P=\hat%
\protect\rho$. The generator $\hat G_{\protect\mu}^{(0)}$ preserves the
projected sector, whereas $\hat G_{\protect\alpha}^{(1)}$ couples the $\hat
P $ and $\hat Q$ sectors. (b) The resulting QFIM and UCM are block diagonal,
with vanishing cross-sector elements $F_{\protect\alpha\protect\mu}^{(0,1)}=0
$ and $\mathcal{I}_{\protect\alpha\protect\mu}^{(0,1)}=0$.}
\label{two_sector}
\end{figure}

In this work, we establish a general symmetry-projection framework that
simultaneously decouples multiparameter sensing channels and enforces weak
compatibility across symmetry sectors. For probe states restricted to a
symmetric subspace, $\hat{\rho}=\hat{P}\hat{\rho}\hat{P}$ via projector $%
\hat{P}$~\cite{Frerot}, encoding generators are partitioned into
subspace-changing ($\hat{G}_{\alpha}^{(1)}$) and subspace-preserving ($\hat{G%
}_{\mu}^{(0)}$) sectors. As depicted in Fig.~\ref{two_sector}, this
structure guarantees an exact sector block-diagonalization: cross-sector
QFIM and UCM elements (or mean SLD commutators) vanish identically,
rendering cross-sector parameters naturally decoupled and weakly compatible.
Furthermore, if projected subspace-changing generators act as scalars in the
occupied subspace ($\hat{P}\hat{G}_{\alpha}^{(1)}\hat{P}\propto\hat{P}$),
the associated QFIM block simplifies strictly to four times the symmetrized
covariance matrix, even for mixed states. This directly grounds
multiparameter sensitivity in experimentally measurable collective
fluctuations.

For parity-protected collective $\mathrm{SU(2)}$ spin systems, the
symmetry-projection mechanism yields a block-diagonal QFIM and eliminates
the transverse--longitudinal UCM element. Under dissipative one-axis
twisting (OAT) with collective dephasing, the generated probe state yields
nearly balanced Heisenberg-scaled sensitivities for $(\theta_y,\theta_z)$
over a broad time window. The mixed steady state retains an isotropic
transverse QFIM, $F_{xx}=F_{yy}\simeq N^2/3$, while transverse compatibility
shows an even--odd parity effect. The sector-resolved mechanism is not
restricted to a two-sector decomposition and can be generalized to
decompositions involving multiple symmetry sectors, where distinct
intersector transitions may support independently estimable and mutually
compatible parameter channels.

We consider the standard framework of multiparameter quantum estimation. For
a state $\hat{\rho}_{\boldsymbol{\theta}}$ with parameters $\boldsymbol{%
\theta}=(\theta_1, \theta_2,\dots)$, the classical Fisher information matrix
(CFIM) $\bm{\mathcal{F}}$ of a given positive operator-valued measure (POVM)
$\{\hat{M}_{x}\}$ is defined by
\begin{equation}
\mathcal{F}_{\alpha \beta }=\sum_{x}\frac{1}{p(x|\boldsymbol{\theta })}%
\left( \frac{\partial p(x|\boldsymbol{\theta })}{\partial \theta _{\alpha }}%
\right) \left( \frac{\partial p(x|\boldsymbol{\theta })}{\partial \theta
_{\beta }}\right) ,  \label{eq:cfim}
\end{equation}%
which quantifies the information extractable from the measurement outcome
probabilities $p(x|\boldsymbol{\theta })=\mathrm{Tr}(\hat{\rho}_{\boldsymbol{%
\theta }}\hat{M}_{x})$. Optimizing over all physically allowable POVMs, the
ultimate precision bound for any unbiased estimator is governed by the QFIM $%
\mathbf{F}$, with entries
\begin{equation}
F_{\alpha \beta }=\frac{1}{2}\mathrm{Tr}\left( \hat{\rho}_{\boldsymbol{%
\theta }}\left\{ \hat{L}_{\alpha },\hat{L}_{\beta }\right\} \right) ,
\label{eq:qfim}
\end{equation}%
where $\hat{L}_{\alpha }$ is the SLD associated with parameter $\theta
_{\alpha }$, defined implicitly via $\partial _{\alpha }\hat{\rho}_{%
\boldsymbol{\theta }}=\frac{1}{2}\{\hat{\rho}_{\boldsymbol{\theta }},\hat{L}%
_{\alpha }\}$. Here, $\partial _{\alpha }\equiv \partial /\partial \theta
_{\alpha }$, $\{\cdot ,\cdot \}$ and $[\cdot ,\cdot ]$ denote the
anticommutator and commutator, respectively. We focus on unitary phase
encoding $\hat{\rho}_{\boldsymbol{\theta }}=e^{-i\boldsymbol{\theta }\cdot
\hat{\mathbf{G}}}\hat{\rho}\,e^{i\boldsymbol{\theta }\cdot \hat{\mathbf{G}}}$%
, generated by a set of Hermitian operators $\hat{\mathbf{G}}=(\hat{G}_{1},%
\hat{G}_{2},\dots )$. At the reference point $\boldsymbol{\theta }=\mathbf{0}
$, the state derivative reduces to $\partial _{\alpha }\hat{\rho}_{%
\boldsymbol{\theta }}|_{\boldsymbol{\theta }=\mathbf{0}}=-i[\hat{G}_{\alpha
},\hat{\rho}]$.

In multiparameter estimation, simultaneous optimality generally requires a
common optimal probe, jointly attainable precision bounds, and statistically
independent parameter estimates~\cite{Ragy2016}. Here we isolate the
encoding-level origin of parameter coupling and show that, for a given probe
state, symmetry alone can enforce the local decoupling of distinct parameter
directions. To establish this result, we restrict the probe to a
symmetry-projected subspace. Let $\hat{P}$ be an orthogonal projector ($\hat{%
P}^{2}=\hat{P}$) such that the support of the probe state is entirely
contained within its range~\cite{Frerot},
\begin{equation}
\hat{P}\hat{\rho}=\hat{\rho}\hat{P}=\hat{\rho}.  \label{eq:probe}
\end{equation}%
We partition the Hermitian generators into subspace-preserving ($\hat{%
\mathbf{G}}^{(0)}$) and subspace-changing ($\hat{\mathbf{G}}^{(1)}$)
sectors. The subspace-preserving generators leave the projected subspace
invariant, whereas the subspace-changing generators couple it to its
orthogonal complement, associated with the projector $\hat{Q}\equiv \mathbb{I%
}-\hat{P}$,
\begin{equation}
\lbrack \hat{G}_{\mu }^{(0)},\hat{P}]=0,\quad \hat{P}\hat{G}_{\alpha }^{(1)}%
\hat{Q}\neq 0,\quad \hat{P}\hat{G}_{\alpha }^{(1)}\hat{P}=g_{\alpha }\hat{P},
\label{eq:generators}
\end{equation}%
with $g_{\alpha }\in \mathbb{R}$. Here, indices $\alpha ,\beta $ label the
subspace-changing sector, whereas $\mu ,\nu $ label the subspace-preserving
sector. The last condition requires each subspace-changing generator to act
as a scalar within the projected sector -- a property we refer to as \textit{%
scalar compression}. For a symmetry-projected probe, this condition further
implies $\langle \hat{G}_{\alpha }^{(1)}\rangle =g_{\alpha }$ and hence
\begin{equation}
\hat{P}\Delta \hat{G}_{\alpha }^{(1)}\hat{P}=0,\qquad \hat{P}\Delta \hat{G}%
_{\alpha }^{(1)}\hat{Q}\neq 0,  \label{eq:scalarcentered}
\end{equation}%
where $\Delta \hat{O}\equiv \hat{O}-\langle \hat{O}\rangle $ and $\langle
\cdot \rangle \equiv \mathrm{Tr}(\hat{\rho}\cdot )$ denotes the expectation
value with respect to the probe state $\hat{\rho}$. Thus, the centered
generators $\Delta \hat{G}_{\alpha }^{(1)}$ act purely off-diagonally on the
projected subspace, coupling it to its orthogonal complement. By contrast, $[%
\hat{G}_{\mu }^{(0)},\hat{P}]=0$ implies $\hat{P}\hat{G}_{\mu }^{(0)}\hat{Q}%
=0$, thereby ensuring that the corresponding state derivatives are fully
supported within the projected subspace.

Although SLDs are non-unique outside the support of a rank-deficient probe $%
\hat{\rho}$, a convenient exact SLD choice on the support of $\hat{\rho}$ is
\begin{equation}
\hat{L}_{\alpha }^{(1)}=-2i[\Delta \hat{G}_{\alpha }^{(1)},\hat{P}]=2i\left(
\hat{P}\Delta \hat{G}_{\alpha }^{(1)}\hat{Q}-\hat{Q}\Delta \hat{G}_{\alpha
}^{(1)}\hat{P}\right) ,  \label{eq:Lperp}
\end{equation}%
which satisfies the defining condition $\frac{1}{2}\{\hat{\rho},\hat{L}%
_{\alpha }^{(1)}\}=-i[\hat{G}_{\alpha }^{(1)},\hat{\rho}]$. For the
subspace-preserving sector, we choose the SLDs to be fully supported within
the projected subspace, while the subspace-changing SLDs constructed above
are purely off-diagonal,
\begin{equation}
\hat{L}_{\mu }^{(0)}=\hat{P}\hat{L}_{\mu }^{(0)}\hat{P},\qquad \hat{P}\hat{L}%
_{\alpha }^{(1)}\hat{P}=\hat{Q}\hat{L}_{\alpha }^{(1)}\hat{Q}=0.
\label{eq:SLDblocks}
\end{equation}%
Equation~\eqref{eq:SLDblocks} implies that the cross-sector products vanish
when sandwiched by the projector $\hat{P}$, i.e., $\hat{P}\hat{L}_{\mu
}^{(0)}\hat{L}_{\alpha }^{(1)}\hat{P}=\hat{P}\hat{L}_{\alpha }^{(1)}\hat{L}%
_{\mu }^{(0)}\hat{P}=0$. Consequently, the cross-sector QFIM elements
identically vanish (see the Supplemental Material~\cite{SM}),
\begin{equation}
F_{\alpha \mu }^{(0,1)}=\frac{1}{2}\mathrm{Tr}\left( \hat{\rho}\left\{ \hat{L%
}_{\alpha }^{(1)},\hat{L}_{\mu }^{(0)}\right\} \right) =0,  \label{eq:Fcross}
\end{equation}%
and the QFIM takes the strictly block-diagonal form
\begin{equation}
\mathbf{F}=%
\begin{pmatrix}
\mathbf{F}^{(1)} & \mathbf{0} \\
\mathbf{0} & \mathbf{F}^{(0)}%
\end{pmatrix}%
.  \label{eq:Fblock}
\end{equation}%
The CFIM obtained from the corresponding SLD-eigenbasis measurements
inherits the same block-diagonal structure as the QFIM~\cite{SM}. This
structure establishes the local decoupling of the two parameter sectors,
preventing Fisher-information cross-talk between subspace-changing and
subspace-preserving channels.

Crucially, the same support argument also eliminates the cross-sector UCM
elements (or mean SLD commutators)~\cite{SM}:
\begin{equation}
\mathcal{I}_{\alpha \mu }^{(0,1)}\equiv \frac{1}{2i}\mathrm{Tr}\left( \hat{%
\rho}\left[ \hat{L}_{\alpha }^{(1)},\hat{L}_{\mu }^{(0)}\right] \right) =0.
\label{eq:weak}
\end{equation}
Hence, the symmetry-sector decomposition induces a joint
block-diagonalization of both the QFIM and UCM. As a result, every
cross-sector parameter pair $(\theta_{\alpha}, \theta_{\mu})$ is strictly
decoupled from information cross-talk and measurement incompatibility, thus
satisfying the weak compatibility condition. This cross-sector feature,
however, does not rule out residual incompatibility within individual
sectors. Furthermore, evaluating the subspace-changing QFIM block using Eq.~%
\eqref{eq:Lperp} yields
\begin{equation}
F_{\alpha \beta }^{(1)}=4\mathrm{Cov}\left( \hat{G}_{\alpha }^{(1)},\hat{G}%
_{\beta }^{(1)}\right) =2\left\langle \left\{ \Delta \hat{G}_{\alpha
}^{(1)},\Delta \hat{G}_{\beta }^{(1)}\right\} \right\rangle ,
\label{eq:Fperp}
\end{equation}%
where $\mathrm{Cov}(\hat{A},\hat{B})\equiv \frac{1}{2}\langle \{\Delta \hat{A%
},\Delta \hat{B}\}\rangle $, $\Delta \hat{O}\equiv \hat{O}-\langle \hat{O}%
\rangle $, and thus $\mathrm{Var}(\hat{O})\equiv\langle(\Delta \hat{O}%
)^2\rangle$. This extends the single-parameter identity of Ref.~\cite{Frerot}
to the symmetrized covariance matrix $\mathbf{\Gamma }^{(1)}$ of an
arbitrary set of subspace-changing generators satisfying the
scalar-compression condition, $\mathbf{F}^{(1)}=4\mathbf{\Gamma }^{(1)}$,
where $(\mathbf{\Gamma }^{(1)})_{\alpha \beta }=\mathrm{Cov}(\hat{G}_{\alpha
}^{(1)},\hat{G}_{\beta }^{(1)})$.

The symmetry-projection framework developed here addresses parameter
independence specifically at the parameter-encoding stage, rather than the
full compatibility problem of multiparameter quantum estimation discussed in
Ref.~\cite{Ragy2016}. For a symmetry-projected probe, separating the phase
generators into subspace-changing and subspace-preserving sectors forces the
cross-sector QFIM and UCM elements vanish simultaneously. This establishes
decoupled and weakly compatible sensing channels across different symmetry
sectors and provides a systematic principle for selecting multiparameter
generators.

We now apply the symmetry-projection theorem to collective $\mathrm{SU}(2)$
systems, where the phase generators are given by $\hat{\mathbf{G}}=\hat{%
\mathbf{J}}=(\hat{J}_{x},\hat{J}_{y}, \hat{J}_{z})$. These systems can be
realized either as an ensemble of $N$ spin-$1/2$ particles (equivalent to a
large spin with $J=N/2$) or as a two-mode bosonic system described within
the Schwinger representation.

We then focus on probe states confined to a definite parity sector, $\hat{%
\rho}=\hat{P}\hat{\rho}\hat{P}$. For a Hermitian parity operator $\hat{\Pi}$
satisfying $\hat{\Pi}^{2}=\mathbb{I}$, we choose $\hat{P}=(\mathbb{I}\pm
\hat{\Pi})/2$ as the projector onto the occupied parity sector, and define $%
\hat{Q}=\mathbb{I}-\hat{P}$ as the projector onto its orthogonal complement,
where
\begin{equation}
\hat{\Pi}=(-1)^{\hat{N}_{\uparrow }}=\exp [i\pi (J+\hat{J}_{z})].
\label{eq:parity}
\end{equation}%
The parity is defined with respect to the chosen spin (or mode) basis $%
\{|\!\uparrow \rangle ,|\!\downarrow \rangle \}$, or equivalently, the $z$
quantization axis. The transverse components $(\hat{J}_{x},\hat{J}_{y})$
flip the parity, $\{\hat{J}_{x,y},\hat{\Pi}\}=0$, and therefore act as
subspace-changing generators. By contrast, the longitudinal component $\hat{J%
}_{z}$ preserves the parity, $[\hat{J}_{z},\hat{\Pi}]=0$, and acts as a
subspace-preserving generator. Applying the symmetry-projection theorem then
yields
\begin{equation}
\mathbf{F}=%
\begin{pmatrix}
\mathbf{F}^{(1)} & \mathbf{0} \\
\mathbf{0} & F_{zz}%
\end{pmatrix}%
=%
\begin{pmatrix}
F_{xx} & F_{xy} & 0 \\
F_{xy} & F_{yy} & 0 \\
0 & 0 & F_{zz}%
\end{pmatrix}%
,  \label{QFIM2}
\end{equation}%
where the transverse ($x,y$) and longitudinal ($z$) sensing sectors are
fully decoupled. Crucially, as the UCM shares this exact block-diagonal
structure, the transverse and longitudinal channels are not only decoupled
at the encoding level but also mutually satisfy the weak-compatibility
condition, specifically evidenced by $F_{xz}=F_{yz}=0$ and $\mathcal{I}_{xz}=%
\mathcal{I}_{yz}=0$ (see also Supplemental Material~\cite{SM}). Furthermore,
parity symmetry enforces $\langle \hat{J}_{x}\rangle =\langle \hat{J}%
_{y}\rangle =0$ and hence $F_{\alpha \beta }=4(\mathbf{\Gamma }%
^{(1)})_{\alpha \beta }=2\langle \{\hat{J}_{\alpha },\hat{J}_{\beta
}\}\rangle $ for $\alpha ,\beta =x$, $y$, where $\mathbf{\Gamma }^{(1)}$ is
a $2\times 2$ covariance matrix.

Notably, the transverse QFIM is fully determined by experimentally
accessible second-order spin correlations, rather than requiring full state
tomography. These correlations can be accessed through collective spin
rotations followed by quantum-nondemolition readout, as demonstrated in
atomic-ensemble experiments~\cite{Xiao2020,jin2024prl,Duan2025,
Zhang2025cooperative}. For a parity-protected probe state, the mean spin
vector is constrained to lie along the $z$ axis, $\langle \hat{\mathbf{J}}%
\rangle =(0,0,\langle \hat{J}_{z}\rangle )$. Consequently, a unit vector in
the transverse plane can be parametrized as ${\mathbf{n}}_{\bot }=(\cos \eta
,\sin \eta )$, and the corresponding transverse spin component is $\hat{J}%
_{\eta }^{(\perp )}=\hat{\mathbf{J}}_{\bot }\cdot {\mathbf{n}}_{\bot }$,
where $\hat{\mathbf{J}}_{\bot }=(\hat{J}_{x},\hat{J}_{y})$ and ${\mathbf{n}}%
_{\bot }\mathbf{n}_{\perp }^{T}=1$. Extremizing the variance $\mathrm{Var}(%
\hat{J}_{\eta }^{(\perp )})=\mathbf{n}_{\perp }\mathbf{\Gamma }^{(1)}\mathbf{%
n}_{\perp }^{T}$ with respect to ${\mathbf{n}}_{\bot }$ yields the extremal
transverse variances~\cite{JinNJP09}:
\begin{equation}
V_{\pm }=\frac{1}{2}\left( \mathcal{C}\pm \sqrt{\mathcal{A}^{2}+\mathcal{B}%
^{2}}\right) ,  \label{V+-}
\end{equation}%
which correspond to the anti-squeezing ($+$) and squeezing ($-$) axes $%
\mathbf{n}_{\pm }$, respectively, and
\begin{equation}
\mathcal{A}=\left\langle \hat{J}_{x}^{2}-\hat{J}_{y}^{2}\right\rangle ,\quad
\mathcal{B}=\left\langle \left\{ \hat{J}_{x},\hat{J}_{y}\right\}
\right\rangle ,\quad \mathcal{C}=\left\langle \hat{J}_{x}^{2}+\hat{J}%
_{y}^{2}\right\rangle .  \label{ABC}
\end{equation}%
The axes $\mathbf{n}_{\pm }$ are the eigenvectors of the $2\times 2$
covariance matrix $\boldsymbol{\Gamma }^{(1)}$, corresponding to the
eigenvalues $V_{\pm }$.

Spin-squeezing experiments typically characterize metrological gain through
the Wineland parameter $\xi_{R}^{2}=NV_{-}/|\langle\hat{\mathbf{J}}%
\rangle|^{2}$~\cite{Wineland1,Wineland1994}, where $V_{-}$ is the minimum
spin variance transverse to the mean spin~\cite{Kitagawa}. Despite its
experimental accessibility, $\xi_{R}^{2}$ becomes ill-defined for vanishing
mean spin and may fail to identify highly entangled probes ~\cite%
{pezze2018quantum,PezzePRL09}. The QFIM remains applicable in this regime,
and its diagonalization identifies the optimal phase-generator directions~%
\cite{Hyllus,Reilly}. For a generic mixed probe, however, constructing the
QFIM generally requires a spectral decomposition of the many-body density
operator. At experimentally relevant sizes---$N\sim10^{5}$ in Bose--Einstein
condensates~\cite{Gross2010,Riedel2010,Mue14,Mao23} and up to $N\sim10^{11}$
in thermal-atom ensembles~\cite%
{Xiao2020,jin2024prl,Duan2025,Zhang2025cooperative}---such a brute-force
construction is generally impractical.

The symmetry-induced block-diagonal structure reduces this problem to
comparing the leading eigenvalues of the transverse and longitudinal
sectors. For the parity-protected probe states considered here, the QFI in
the transverse sector reads $F_{Q}^{(\perp )}=4\mathrm{Var}(\hat{J}_{\eta
}^{(\perp )})=4\mathbf{n}_{\perp }\mathbf{\Gamma }^{(1)}\mathbf{n}_{\perp
}^{T}$, which gives the largest eigenvalues $4V_{+}$, as Eq.~\eqref{V+-}.
Comparing it with $F_{zz}$ yields
\begin{equation}
F_{Q,\max }=\max \{4V_{+},F_{zz}\},\qquad \mathbf{n}_{\max }\in \{\mathbf{n}%
_{+},\hat{\mathbf{z}}\},
\end{equation}%
where $V_{+}$ and $\mathbf{n}_{+}$ correspond to the anti-squeezed variance
and its direction. When $4V_{+}=F_{zz}$, any normalized direction in $%
\mathrm{span}\{\mathbf{n}_{+},\hat{\mathbf{z}}\}$ is optimal.

The above result does not depend on choosing $\hat{\mathbf{z}}$ as the
parity axis. For parity about an arbitrary axis $\mathbf{n}$, the QFIM
separates into longitudinal and transverse sectors relative to $\mathbf{n}$.
When the nonzero mean spin is aligned with $\mathbf{n}$, the directions $%
\mathbf{n}$ and $\mathbf{n}_{+}$ maximize the longitudinal and transverse
QFI, respectively. The optimal single-parameter direction can therefore
switch between sectors only when their leading eigenvalues cross. This
analytical sector-selection rule makes explicit the mechanism underlying
previous numerical observations~\cite{Reilly}.

Parity-protected probes can be generated by the particle-number-conserving
master equation
\begin{equation}
\frac{\partial\hat{\rho}}{\partial t} =-i[\hat{H},\hat{\rho}] +\frac{\gamma}{%
2} \left(2\hat{J}_{z}\hat{\rho}\hat{J}_{z} -\{\hat{J}_{z}^{2},\hat{\rho}%
\}\right),  \label{master}
\end{equation}
where $\hbar=1$. If both the coherent and dissipative dynamics preserve
parity, $[\hat{H},\hat{\Pi}]=[\hat{J}_{z},\hat{\Pi}]=0$, the Liouvillian
does not couple different parity sectors. An initial state supported
entirely in one sector therefore remains that sector, i.e., $\hat{\rho}(t)=%
\hat{P}\hat{\rho}(t)\hat{P}$ at any $t$, despite the loss of purity induced
by dephasing. As a concrete example, we consider the OAT Hamiltonian $\hat{H}%
_{\mathrm{OAT}}=\chi\hat{J}_{x}^{2}$ acting on the even-parity initial state
$|J,-J\rangle$, a setting proposed for generating a GHZ state ~\cite%
{MolmerPRL99}. The evolution remains within the even-parity sector. This
scheme readily accommodates other parity-preserving collective interactions,
including two-axis twisting (TAT)~\cite%
{Kitagawa,liu2011prl,yukawa2014pra,Kajtoch,Manuel} and collective XYZ models~%
\cite{YCLiuPRL24,miller2024,luo2025}.

\begin{figure}[hptb]
\centering
\includegraphics[width=\columnwidth]{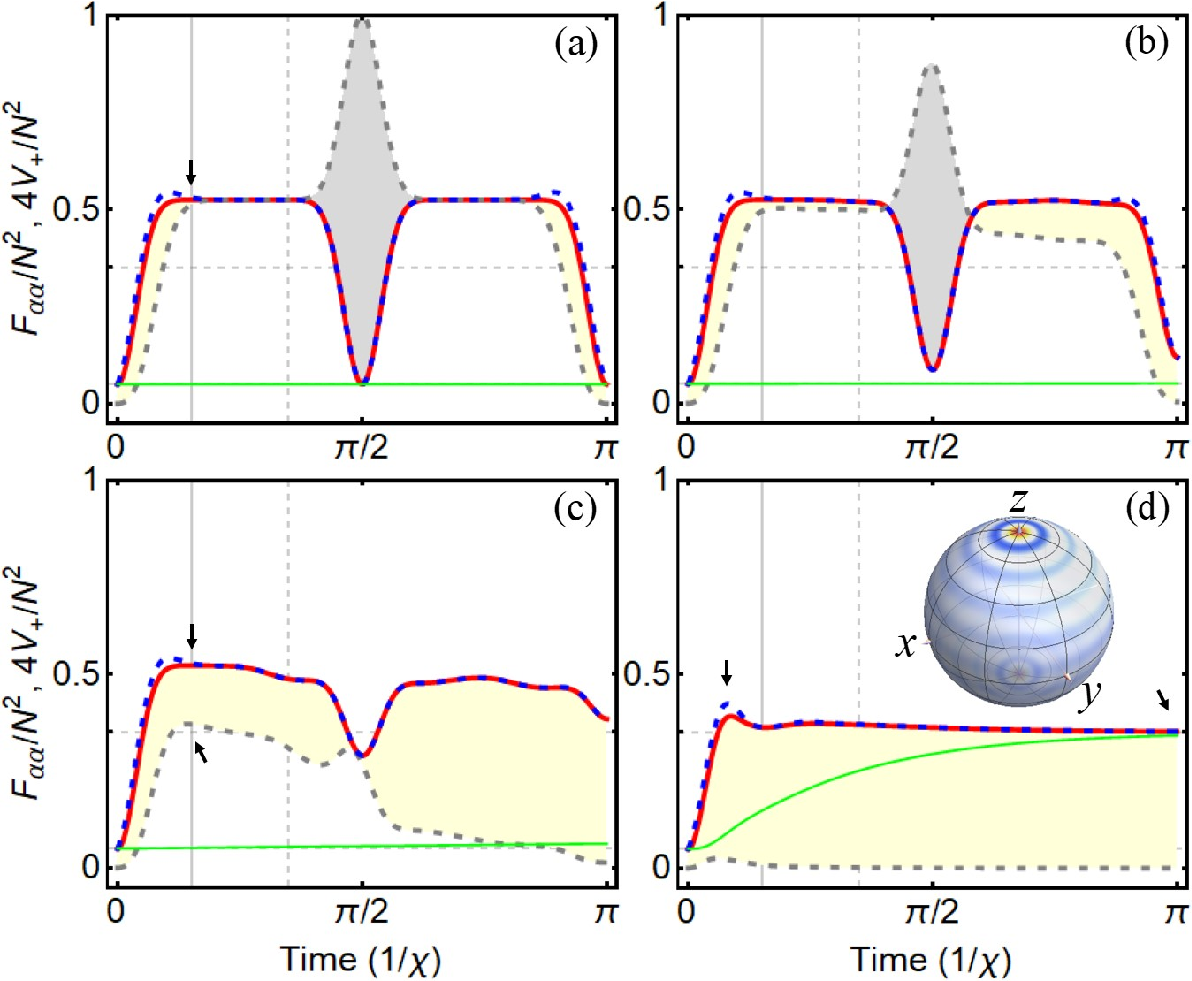}
\caption{Time evolution of the maximal QFI $4V_{+}$ (blue dashed) and the
QFIM elements $F_{xx}$ (green solid), $F_{yy}$ (red solid), and $F_{zz}$
(gray dashed) under dissipative OAT dynamics for (a) $\protect\gamma/\protect%
\chi=0$, (b) $10^{-3}$, (c) $10^{-2}$, and (d) $10^{0}$. Under weak
dephasing, a broad plateau develops within $3\protect\tau_{d}\lesssim\protect%
\chi t\lesssim\protect\pi/2-3\protect\tau_{d}$, where $4V_{+}\simeq
F_{yy}\approx F_{zz}\approx N^2/2$ with $\protect\tau_{d}=1/\protect\sqrt{2N}
$. At $\protect\gamma/\protect\chi=10^{-2}$, the transient QFIs remain
sizable, reaching $F_{yy}\sim0.5N^{2}$ and $F_{zz}\sim0.3N^{2}$ (indicated
by arrows). For strong dephasing ($\protect\gamma\sim\protect\chi$), $F_{zz}$
is rapidly suppressed, whereas the transverse QFI $F_{yy}$ near $\protect%
\chi t\sim1/\protect\sqrt{N}$ remains slightly above the steady-state QFI $%
F_{Q,\mathrm{ss}}\simeq N^{2}/3$ (horizontal dotted line). Inset in (d):
steady-state quasiprobability distribution. All panels use $N=20$.}
\label{fig2}
\end{figure}

Figure~\ref{fig2} illustrates the two-parameter sensing scheme based on the
OAT dynamically evolved state. The optimal sensing directions, $\mathbf{n}%
_{+}\simeq\hat{\mathbf{y}}$ and $\mathbf{n}=\hat{\mathbf{z}}$, yield fully
diagonal QFIM and UCM for $(\theta_y,\theta_z)$, which guarantees complete
decoupling of information cross-talk and weakly compatibility. In the
unitary limit ($\gamma=0$), the OAT dynamics exhibits an extended plateau~%
\cite{PezzePRL09}, during which $4V_{+}\simeq F_{yy}\simeq F_{zz}\simeq
N(N+1)/2$, within the time window $3\tau_d\lesssim\chi
t\lesssim\pi/2-3\tau_d $, where $\tau_d=1/\sqrt{2N}$ (see Ref.~\cite%
{PezzePRL09}, and the Supplemental Material~\cite{SM}). The parameters $%
(\theta_y,\theta_z)$ can therefore be estimated through two nearly balanced
channels with $N^2$ scaling. Under weak-to-moderate dephasing, both $F_{yy}$
and $F_{zz}$ simultaneously retain the $N^2$ scaling over a broad time
window, $1/\sqrt{N}\lesssim\chi t\lesssim\pi/2$ [Figs.~\ref{fig2}(b) and \ref%
{fig2}(c)]. Dissipative OAT therefore enables weakly compatible,
simultaneous estimation of $(\theta_y,\theta_z)$ using a single collective
probe, without additional degrees of freedom or spatial partitioning.

Although parity enforces $F_{yz}=\mathcal{I}_{yz}=0$ for any dephasing
strength, strong dephasing drives $F_{zz}$ to zero [Fig.~\ref{fig2}(d)] and
thereby destroys the simultaneous transverse--longitudinal enhancement as
the system approaches the maximally mixed state within the even-parity
sector,
\begin{equation}
\hat{\rho}_{\mathrm{ss}} = \frac{\hat{P}}{d_N} = \frac{1}{d_N} \sum_{m\in%
\mathcal{M}_+} |J,m\rangle\langle J,m|,  \label{eq:steady_state}
\end{equation}
where $d_N=\lceil(N+1)/2\rceil$ denotes the dimension of the even-parity
subspace, with $\mathcal{M}_+=\{-J,-J+2,\ldots,J\}$ for even $N$, and $%
\mathcal{M}_+=\{-J,-J+2,\ldots,J-1\}$ for odd $N$. Its purity is given by $%
\mathrm{Tr}(\hat{\rho}_{\mathrm{ss}}^2)=1/d_N$. Because $[\hat{\rho}_{%
\mathrm{ss}},\hat{J}_z]=0$, the longitudinal QFI vanishes ($F_{zz}=0$),
whereas combined parity and transverse rotational symmetry yield $\mathbf{F}%
_{\mathrm{ss}} = F_{Q,\mathrm{ss}}\mathrm{diag}(1,1,0)$, with
\begin{equation}
F_{Q,\mathrm{ss}} =
\begin{cases}
N(N+1)/3, & \text{(even $N$)}, \\[2pt]
N(N+2)/3, & \text{(odd $N$)}.%
\end{cases}
\label{steadyQFI}
\end{equation}
At long times, $F_{xx}$ and $F_{yy}$ converge to the same steady-state
value, consistent with the isotropic quasiprobability distribution shown in
Fig.~\ref{fig2}(d). Consequently, any sensing direction in the $xy$ plane
yields $F_{Q,\mathrm{ss}}\simeq N^2/3$ for large $N$, despite the highly
mixed nature of the steady state.

The transverse isotropy of the steady-state QFIM, $F_{xx}=F_{yy}\simeq
N^{2}/3$, naturally suggests a two-parameter encoding generated by $(\hat{J}%
_{x},\hat{J}_{y})$. Isotropy alone, however, does not guarantee the
simultaneous attainability of the corresponding QCRBs. A related scheme was
realized in a spin-$1$ BEC experiment~\cite{Cao2025}, where zero
magnetization ensures weak compatibility between two parameters encoded by
noncommuting transverse rotations. For the dissipative-OAT steady state,
however, $\mathcal{I}_{xy} =2\langle\hat{J}_{z}\rangle_{\mathrm{ss}}$, so
compatibility depends sharply on particle-number parity---failing for odd $N$
but restored for even $N$~\cite{SM}. By contrast, the parity symmetry
guarantees channel decoupling and weak compatibility for the
transverse--longitudinal encoding, irrespective of whether $N$ is even or
odd.

Although the longitudinal QFIM component satisfies $F_{zz}\leq 4\mathrm{Var}%
(\hat J_z)$ and generally requires a full mixed-state QFIM evaluation based
on a spectral decomposition, the two-sector framework developed here remains
broadly applicable to parity-conserving systems. For the dissipative TAT
model, this construction resolves the weak-incompatibility issue associated
with the transverse $(\theta_x,\theta_y)$ encoding, while rendering the
enhanced transverse components, $F_{xx}=F_{yy}\simeq N^2/3$, directly
accessible without density-matrix diagonalization. A fuller account of this
dissipative sensing phenomenology is reserved for future work, where
designing feasible joint measurements that saturate the corresponding
mixed-state multiparameter bounds remains a central challenge~\cite%
{Pezze2017,Matsumoto2002,WangPRL2026}.

In summary, we have established symmetry projection as a general organizing
principle for multiparameter quantum sensing, yielding decoupled sensing
sectors with weak compatibility. For parity-protected collective $\mathrm{SU}%
(2)$ probes, the framework relates transverse QFIMs to measurable spin
covariances and reveals Heisenberg-scaled transverse--longitudinal
sensitivities in dissipative OAT dynamics, together with an isotropic
transverse $N^{2}/3$ scaling at steady state. Beyond the minimal two-sector
realization, the underlying sector-resolved structure provides a route
toward compatible multiparameter sensing in more general symmetry-protected
quantum systems~\cite{SM}. These results demonstrate how symmetry-resolved structures
can enable weakly compatible multiparameter metrology in open quantum
many-body systems.

\textit{Acknowledgments.---}Project supported by Quantum Science and
Technology-National Science and Technology Major Project (Grant No.
2023ZD0300904), the National Natural Science Foundation of China (NSFC)
Grant No.~12405028 and No.~12274019, and the NSAF grant in NSFC with grant
No.~U2230402. We acknowledge the Zhejiang Provincial Natural Science
Foundation of China under Grant No.~LQ24A050002, and the computational
support from the Beijing Computational Science Research Center (CSRC).

% ===================================================================
% 主手稿正文结束
% ===================================================================

\onecolumngrid % <--- 核心命令：强制从这里开始打破双栏，进入全栏通栏排版
\clearpage

\setcounter{equation}{0} \setcounter{figure}{0} \renewcommand{%
\theequation}{S\arabic{equation}} \renewcommand{\thefigure}{S\arabic{figure}}

\section*{Supplemental Material for ``Symmetry-Projected Weakly Compatible Multiparameter Quantum Sensing''}

This Supplemental Material provides explicit analytical derivations and extended theoretical results supporting the symmetry-projection framework presented in the main text. The contents are organized into two main sections:

\begin{enumerate}

\item \emph{Symmetry-Projected Theorem for Multiparameter Quantum Sensing.}
We establish the algebraic framework for symmetry-projected probe states through explicit constructions of symmetric logarithmic derivatives (SLDs):
\begin{itemize}
\item \textbf{Two-sector symmetry-projection theorem:} Proof of the common block-diagonal structure induced on both the quantum Fisher information matrix (QFIM) and the Uhlmann curvature matrix (UCM), along with the exact reduction of the QFIM to the symmetrized covariance matrix;
\item \textbf{Projective measurements and classical Fisher information:} Derivation of the classical Fisher information matrix (CFIM) under SLD eigenbasis measurements, showing that measurement-level information decoupling directly inherits the QFIM symmetry;
\item \textbf{Three-sector extension:} Generalization of the SLD constructions and block-diagonal QFIM/UCM structures to systems with multiple orthogonal symmetry sectors.
\end{itemize}

\item \emph{Application to Parity-Protected Collective $\mathrm{SU}(2)$ Systems.}
We apply the framework to parity-symmetric spin systems and explore optimal multiparameter phase estimation:
\begin{itemize}
\item \textbf{Spectral decomposition and parity selection rules:} Independent re-derivation of the diagonal QFIM structure and identification of optimal sensing axes defined by transverse anti-squeezed quadratures and longitudinal mean-spin directions;
\item \textbf{UCM evaluation and weak compatibility:} Direct SLD-based derivation of UCM elements, analysing weak compatibility for both transverse--transverse and transverse--longitudinal parameter pairs;
\item \textbf{Exact results for one-axis twisting (OAT) states:} Analytical characterization of pure OAT probe states, revealing the explicit dependence of optimal sensing axis switching on the particle-number parity of $N$.
\end{itemize}

\end{enumerate}

\subsection*{Symmetry-Projected theorem for multiparameter quantum sensing}

\subsubsection*{I. Proof via symmetric logarithmic derivatives (SLDs)}

We consider a generic probe state $\hat{\rho}$ fully supported within a
subspace of the Hilbert space~\cite{Frerot_supp},
\begin{equation}
\hat{\rho}=\hat{P}\hat{\rho}=\hat{\rho}\hat{P}=\hat{P}\hat{\rho}\hat{P},
\label{smeq:rho}
\end{equation}%
where $\hat{P}$ is an orthogonal projector ($\hat{P}^{2}=\hat{P}$) and $\hat{%
Q}\equiv \mathbb{I}-\hat{P}$. For a unitary phase-encoding state $\hat{\rho}%
_{\boldsymbol{\theta }}=e^{-i\boldsymbol{\theta }\cdot \hat{\mathbf{G}}}\hat{%
\rho}e^{i\boldsymbol{\theta }\cdot \hat{\mathbf{G}}}$, we partition the
Hermitian generators $\hat{\mathbf{G}}=(\hat{G}_{1},\hat{G}_{2},\dots )$
into transverse and longitudinal sectors, as Eq.~(4) in main text. The
longitudinal generators ($\hat{\in \mathbf{G}}^{(0)}$) leave the projected
subspace invariant, $[\hat{G}_{\mu }^{(0)},\hat{P}]=0$, whereas the
transverse generators ($\hat{\mathbf{G}}^{(1)}$) connect it to its
orthogonal complement, i.e., $\hat{P}\hat{G}_{\alpha }^{(1)}\hat{Q}\neq 0$,
and satisfy the scalar compression:
\begin{equation}
\hat{P}\hat{G}_{\alpha }^{(1)}\hat{P}=g_{\alpha }\hat{P},\qquad (g_{\alpha
}\in \mathbb{R}).
\end{equation}%
Here we use the indices $\alpha ,\beta $ and $\mu ,\nu $ label transverse
and longitudinal parameters, respectively.

Using Eq.~\eqref{smeq:rho} and the cyclicity of the trace, we obtain
\begin{equation}
\left\langle \hat{G}_{\alpha }^{(1)}\right\rangle =\mathrm{Tr}\left( \hat{%
\rho}\hat{P}\hat{G}_{\alpha }^{(1)}\hat{P}\right) =g_{\alpha }\mathrm{Tr}%
\left( \hat{\rho}\hat{P}\right) =g_{\alpha },
\end{equation}%
where $\mathrm{Tr}(\hat{\rho}\hat{P})=\mathrm{Tr}(\hat{\rho})=1$.
Consequently, the centered transverse generators $\Delta \hat{G}_{\alpha
}^{(1)}=\hat{G}_{\alpha }^{(1)}-\langle \hat{G}_{\alpha }^{(1)}\rangle $
vanish under projection
\begin{equation}
\hat{P}\Delta \hat{G}_{\alpha }^{(1)}\hat{P}=0.  \label{smeq:scale}
\end{equation}%
Since $\Delta \hat{G}_{\alpha }^{(1)}$ connect $\hat{P}$ to its orthogonal
complement $\hat{Q}$, we choose the transverse SLDs as
\begin{equation}
\hat{L}_{\alpha }^{(1)}=-2i\left[ \Delta \hat{G}_{\alpha }^{(1)},\hat{P}%
\right] =2i\left( \hat{P}\Delta \hat{G}_{\alpha }^{(1)}\hat{Q}-\hat{Q}\Delta
\hat{G}_{\alpha }^{(1)}\hat{P}\right) .  \label{smeq:SLD1}
\end{equation}%
Using Eq.~\eqref{smeq:rho}, we verify%
\begin{equation}
\left. \partial _{\alpha }\hat{\rho}_{\boldsymbol{\theta }}\right\vert _{%
\boldsymbol{\theta }=0}=\frac{1}{2}\left\{ \hat{\rho},\hat{L}_{\alpha
}^{(1)}\right\} =-i\left[ \Delta \hat{G}_{\alpha }^{(1)},\hat{\rho}\right]
=-i\left[ \hat{G}_{\alpha }^{(1)},\hat{\rho}\right] ,
\end{equation}%
where $\partial _{\alpha }=\partial /\partial \theta _{\alpha }$.
Furthermore, we obtain
\begin{equation}
\mathrm{Tr}\left( \hat{\rho}\hat{L}_{\alpha }^{(1)}\hat{L}_{\beta
}^{(1)}\right) =4\mathrm{Tr}\left( \hat{\rho}\hat{P}\Delta \hat{G}_{\alpha
}^{(1)}\Delta \hat{G}_{\beta }^{(1)}\hat{P}\right) =4\left\langle \Delta
\hat{G}_{\alpha }^{(1)}\Delta \hat{G}_{\beta }^{(1)}\right\rangle ,
\end{equation}%
which is symmetric with respect to $\alpha $ and $\beta $. Thus, the
transverse QFIM block reduces to four times the symmetrized covariance
matrix,
\begin{equation}
F_{\alpha \beta }^{(1)}=\frac{1}{2}\mathrm{Tr}\left( \hat{\rho}\left\{ \hat{L%
}_{\alpha }^{(1)},\hat{L}_{\beta }^{(1)}\right\} \right) =4\mathrm{Cov}%
\left( \hat{G}_{\alpha }^{(1)},\hat{G}_{\beta }^{(1)}\right) ,
\end{equation}%
where $\mathrm{Cov}(\hat{A},\hat{B})\equiv \frac{1}{2}\langle \{\Delta \hat{A%
},\Delta \hat{B}\}\rangle $, with $\Delta \hat{O}\equiv \hat{O}-\langle \hat{%
O}\rangle $.

For the longitudinal sector, a valid choice of SLDs is
\begin{equation}
\hat{L}_{\mu }^{(0)}=\hat{P}\hat{L}_{\mu }^{(0)}\hat{P},  \label{eq:SLD2}
\end{equation}%
which are compatible with $[\hat{G}_{\mu }^{(0)},\hat{P}]=0$ and hence $\hat{%
P}\hat{G}_{\mu }^{(0)}\hat{Q}=0$. Since $\hat{P}\hat{L}_{\alpha }^{(1)}\hat{P%
}=0$ and $\hat{P}\hat{L}_{\mu }^{(0)}\hat{Q}=\hat{Q}\hat{L}_{\mu }^{(0)}\hat{%
P}=0$, we have
\begin{equation}
\hat{P}\hat{L}_{\alpha }^{(1)}\hat{L}_{\mu }^{(0)}\hat{P}=\hat{P}\hat{L}%
_{\alpha }^{(1)}\left( \hat{P}+\hat{Q}\right) \hat{L}_{\mu }^{(0)}\hat{P}%
=\left( \hat{P}\hat{L}_{\alpha }^{(1)}\hat{P}\right) \hat{L}_{\mu }^{(0)}%
\hat{P}+\hat{P}\hat{L}_{\alpha }^{(0)}\left( \hat{Q}\hat{L}_{\mu }^{(1)}\hat{%
P}\right) =0,
\end{equation}%
and thus%
\begin{equation}
\mathrm{Tr}\left( \hat{\rho}\hat{L}_{\alpha }^{(1)}\hat{L}_{\mu
}^{(0)}\right) =\mathrm{Tr}\left( \hat{P}\hat{\rho}\hat{P}\hat{L}_{\alpha
}^{(1)}\hat{L}_{\mu }^{(0)}\right) =\mathrm{Tr}\left( \hat{\rho}\hat{P}\hat{L%
}_{\alpha }^{(1)}\hat{L}_{\mu }^{(0)}\hat{P}\right) =0.
\end{equation}%
Therefore, the cross-sector QFIM elements identically vanish,
\begin{equation}
F_{\alpha \mu }^{(0,1)}=\frac{1}{2}\mathrm{Tr}\left( \hat{\rho}\left\{ \hat{L%
}_{\alpha }^{(1)},\hat{L}_{\mu }^{(0)}\right\} \right) =0,
\label{eq:TLcross}
\end{equation}%
leading to the block-diagonal QFIM, as Eq.~(8) in main text. Moreover, the
weak compatibility condition for a pair transverse-longitudinal parameters $%
(\theta _{\alpha },\theta _{\mu })$ is automatically fulfilled,
\begin{equation}
\mathcal{I}_{\alpha \mu }^{(0,1)}\equiv \frac{1}{2i}\mathrm{Tr}\left( \hat{%
\rho}\left[ \hat{L}_{\alpha }^{(1)},\hat{L}_{\mu }^{(0)}\right] \right) =0.
\label{compatibility}
\end{equation}%
As illustrated in Fig.~1 of the main text,, the QFIM and the UCM then have
the direct-sum forms
\begin{equation}
\mathbf{F}=\mathbf{F}^{(0)}\oplus \mathbf{F}^{(1)},\qquad \boldsymbol{%
\mathcal{I}}=\boldsymbol{\mathcal{I}}^{(0)}\oplus \boldsymbol{\mathcal{I}}%
^{(1)},  \label{eq:two-block-matrices}
\end{equation}%
or equivalently,
\begin{equation}
\mathbf{F}=\left(
\begin{array}{cc}
\mathbf{F}^{(0)} & \mathbf{0} \\
\mathbf{0} & \mathbf{F}^{(1)}%
\end{array}%
\right) ,\qquad \boldsymbol{\mathcal{I}}=\left(
\begin{array}{cc}
\boldsymbol{\mathcal{I}}^{(0)} & \mathbf{0} \\
\mathbf{0} & \boldsymbol{\mathcal{I}}^{(1)}%
\end{array}%
\right) .  \label{eq:two-explicit-block-matrices}
\end{equation}

\subsubsection*{II. The CFIM of projective SLD-eigebasis measurements}

We now consider projective measurements in the eigenbasis of the respective
SLD operators and calculate the CFIM. \textit{Transverse Sector CFIM.---}Let
$\{|k\rangle _{\alpha }\}$ be the eigenbasis of $\hat{L}_{\alpha }^{(1)}$:
\begin{equation}
\hat{L}_{\alpha }^{(1)}|k\rangle _{\alpha }=\lambda _{k}^{(\alpha
)}|k\rangle _{\alpha },\qquad \left( \lambda _{k}^{(\alpha )}\in \mathbb{R}%
\right) ,  \label{smeq:Laeigenbasis}
\end{equation}%
and define the conditional probability distribution:
\begin{equation}
p_{k}^{(\alpha )}\equiv p_{\alpha }(k|\boldsymbol{\theta })=\mathrm{Tr}%
\left( \hat{\rho}_{\boldsymbol{\theta }}|k\rangle _{\alpha }\langle
k|_{\alpha }\right) =\langle k|_{\alpha }\hat{\rho}_{\boldsymbol{\theta }%
}|k\rangle _{\alpha }.
\end{equation}%
Before details, we introduce the spectral decomposition $\hat{L}_{\alpha
}^{(1)}=\sum_{k}\lambda _{k}^{(\alpha )}|k\rangle _{\alpha }\langle
k|_{\alpha }$ and two useful identities
\begin{eqnarray}
\sum_{k}\lambda _{k}^{(\alpha )}\langle k|_{\alpha }\hat{A}|k\rangle
_{\alpha } &=&\sum_{k}\langle k|_{\alpha }\hat{L}_{\alpha }^{(1)}\hat{A}%
|k\rangle _{\alpha }=\mathrm{Tr}\left( \hat{A}\hat{L}_{\alpha }^{(1)}\right)
,\qquad   \label{smeq;iden1} \\
\sum_{k}\left( \lambda _{k}^{(\alpha )}\right) ^{2}\langle k|_{\alpha }\hat{A%
}|k\rangle _{\alpha } &=&\sum_{k}\langle k|_{\alpha }\left( \hat{L}_{\alpha
}^{(1)}\right) ^{2}\hat{A}|k\rangle _{\alpha }=\mathrm{Tr}\left( \hat{A}%
\left( \hat{L}_{\alpha }^{(1)}\right) ^{2}\right) ,  \label{smeq:iden2}
\end{eqnarray}%
which holds for any operator $\hat{A}$.

Using the SLD equation $\partial \hat{\rho}_{\boldsymbol{\theta }}/\partial
\theta _{\alpha }=\frac{1}{2}(\hat{\rho}_{\boldsymbol{\theta }}\hat{L}%
_{\alpha }^{(1)}+\hat{L}_{\alpha }^{(1)}\hat{\rho}_{\boldsymbol{\theta }})$,
we have
\begin{align}
\frac{\partial p_{k}^{(\alpha )}}{\partial \theta _{\alpha }}& =\mathrm{Tr}%
\left( \frac{\partial \hat{\rho}_{\boldsymbol{\theta }}}{\partial \theta
_{\alpha }}|k\rangle _{\alpha }\langle k|_{\alpha }\right)
=\sum_{l}\left\langle l\right\vert _{\alpha }\left( \frac{\partial \hat{\rho}%
_{\boldsymbol{\theta }}}{\partial \theta _{\alpha }}|k\rangle _{\alpha
}\langle k|_{\alpha }\right) \left\vert l\right\rangle _{\alpha }  \notag \\
& =\frac{1}{2}\left\langle k\right\vert _{\alpha }\left( \hat{\rho}_{%
\boldsymbol{\theta }}\hat{L}_{\alpha }^{(1)}+\hat{L}_{\alpha }^{(1)}\hat{\rho%
}_{\boldsymbol{\theta }}\right) |k\rangle _{\alpha },  \label{smeq:dpk1} \\
& =\lambda _{k}^{(\alpha )}\langle k|_{\alpha }\hat{\rho}_{\boldsymbol{%
\theta }}|k\rangle _{\alpha }=\lambda _{k}^{(\alpha )}p_{k}^{(\alpha )}.
\label{smeq:dpk2}
\end{align}%
Therefore, we obtain transverse diagonal CFIM elements,
\begin{align}
\mathcal{F}_{\alpha \alpha }^{(1)}& =\sum_{k}\frac{1}{p_{k}^{(\alpha )}}%
\left( \frac{\partial p_{k}^{(\alpha )}}{\partial \theta _{\alpha }}\right)
^{2} =\sum_{k}p_{k}^{(\alpha )}(\lambda _{k}^{(\alpha
)})^{2}\notag \\
&=\sum_{k}(\lambda _{k}^{(\alpha )})^{2}\langle k|_{\alpha }\hat{\rho}%
_{\boldsymbol{\theta }}|k\rangle _{\alpha }   =\mathrm{Tr}\left[ \hat{\rho}_{\boldsymbol{\theta }}(\hat{L}_{\alpha
}^{(1)})^{2}\right] \equiv F_{\alpha \alpha }^{(1)}.  \label{smeq:Faa}
\end{align}%
where, in the last step, we use Eq.~\eqref{smeq:iden2}. Similarly, we obtain
the transverse off-diagonal elements
\begin{align}
\mathcal{F}_{\alpha \beta }^{(1)}& =\sum_{k}\frac{1}{p_{k}^{(\alpha )}}%
\left( \frac{\partial p_{k}^{(\alpha )}}{\partial \theta _{\alpha }}\right)
\left( \frac{\partial p_{k}^{(\alpha )}}{\partial \theta _{\beta }}\right)
=\sum_{k}\lambda _{k}^{(\alpha )}\frac{\partial p_{k}^{(\alpha )}}{\partial
\theta _{\beta }}=\frac{1}{2}\sum_{k}\lambda _{k}^{(\alpha )}\langle
k|_{\alpha }\left( \hat{\rho}_{\boldsymbol{\theta }}\hat{L}_{\beta }^{(1)}+%
\hat{L}_{\beta }^{(1)}\hat{\rho}_{\boldsymbol{\theta }}\right) |k\rangle
_{\alpha }  \notag \\
& =\frac{1}{2}\mathrm{Tr}\left[ \hat{L}_{\alpha }^{(1)}\left( \hat{\rho}_{%
\boldsymbol{\theta }}\hat{L}_{\beta }^{(1)}+\hat{L}_{\beta }^{(1)}\hat{\rho}%
_{\boldsymbol{\theta }}\right) \right] =\frac{1}{2}\mathrm{Tr}\left[ \hat{%
\rho}_{\boldsymbol{\theta }}\left( \hat{L}_{\beta }^{(1)}\hat{L}_{\alpha
}^{(1)}+\hat{L}_{\alpha }^{(1)}\hat{L}_{\beta }^{(1)}\right) \right] \equiv
F_{\alpha \beta }^{(1)}.  \label{smeq:Fab}
\end{align}%
where we used Eqs.~\eqref{smeq:dpk2}, \eqref{smeq:dpk1}, and %
\eqref{smeq;iden1}, as well as the cyclicity of the trace. Thus, the
transverse block CFIM $\bm{\mathcal{F}}^{(1)}=\mathbf{F}^{(1)}$.

\textit{Cross-Sector CFIM.---}For any transverse parameter $\alpha $ and
longitudinal parameter $\mu $, the CFIM elements are defined as
\begin{equation}
\mathcal{F}_{\alpha \mu }^{(0,1)}=\sum_{k}\frac{1}{p_{k}^{(\alpha )}}\left(
\frac{\partial p_{k}^{(\alpha )}}{\partial \theta _{\alpha }}\right) \left(
\frac{\partial p_{k}^{(\alpha )}}{\partial \theta _{\mu }}\right) .
\end{equation}%
We measure in the transverse SLD eigenbasis $\{|k\rangle _{\alpha }\}$, so $%
\partial p_{k}^{(\alpha )}/\partial \theta _{\alpha }=\lambda _{k}^{(\alpha
)}p_{k}^{(\alpha )}$, as Eq.~\eqref{smeq:dpk2}, and
\begin{equation}
\frac{\partial p_{k}^{(\alpha )}}{\partial \theta _{\mu }}=\mathrm{Tr}\left(
\frac{\partial \hat{\rho}_{\boldsymbol{\theta }}}{\partial \theta _{\mu }}%
|k\rangle _{\alpha }\langle k|_{\alpha }\right) =\frac{1}{2}\left\langle
k\right\vert _{\alpha }\left( \hat{\rho}_{\boldsymbol{\theta }}\hat{L}_{\mu
}^{(0)}+\hat{L}_{\mu }^{(0)}\hat{\rho}_{\boldsymbol{\theta }}\right)
|k\rangle _{\alpha }.
\end{equation}%
Therefore, we have%
\begin{align}
\mathcal{F}_{\alpha \mu }^{(0,1)}& =\sum_{k}\lambda _{k}^{(\alpha )}\frac{%
\partial p_{k}^{(\alpha )}}{\partial \theta _{\mu }} =\frac{1}{2}%
\sum_{k}\lambda _{k}^{(\alpha )}\langle k|_{\alpha }\left( \hat{\rho}_{%
\boldsymbol{\theta }}\hat{L}_{\mu }^{(0)}+\hat{L}_{\mu }^{(0)}\hat{\rho}_{%
\boldsymbol{\theta }}\right) |k\rangle _{\alpha }  \notag \\
& =\frac{1}{2}\mathrm{Tr}\left[ \hat{L}_{\alpha }^{(1)}\left( \hat{\rho}_{%
\boldsymbol{\theta }}\hat{L}_{\mu }^{(0)}+\hat{L}_{\mu }^{(0)}\hat{\rho}_{%
\boldsymbol{\theta }}\right) \right] =\frac{1}{2}\mathrm{Tr}\left[ \hat{\rho}%
_{\boldsymbol{\theta }}\left( \hat{L}_{\mu }^{(0)}\hat{L}_{\alpha }^{(1)}+%
\hat{L}_{\alpha }^{(1)}\hat{L}_{\mu }^{(0)}\right) \right] \equiv F_{\alpha
\mu }^{(0,1)},  \label{smeq:Famu}
\end{align}%
where we used the cyclicity of the trace. Since $F_{\alpha \mu }^{(0,1)}=0$,
as Eq.~\eqref{eq:TLcross}, we further obtain
\begin{equation}
\mathcal{F}_{\alpha \mu }^{(0,1)}=F_{\alpha \mu }^{(0,1)}=0.
\end{equation}

\textit{Longitudinal Sector CFIM.---}Let $\{|m\rangle _{\mu }\}$ be the
eigenbasis of $\hat{L}_{\mu }^{(0)}$:
\begin{equation}
\hat{L}_{\mu }^{(0)}|m\rangle _{\mu }=\lambda _{m}^{(\mu )}|m\rangle _{\mu
},\qquad \left( \lambda _{m}^{(\mu )}\in \mathbb{R}\right) .
\end{equation}%
Since $\hat{L}_{\mu }^{(0)}=\hat{P}\hat{L}_{\mu }^{(0)}\hat{P}$, all
eigenstates lie in $\hat{P}$ and thus $\hat{P}|m\rangle _{\mu }=|m\rangle
_{\mu }$. Define the outcome probabilities
\begin{equation}
p_{m}^{(\mu )}=p_{\mu }(m|\boldsymbol{\theta })=\mathrm{Tr}\left( \hat{\rho}%
_{\boldsymbol{\theta }}|m\rangle _{\mu }\langle m|_{\mu }\right) .
\end{equation}%
Completely analogously to the transverse case, we have
\begin{equation}
\mathcal{F}_{\mu \nu }^{(0)}=\frac{1}{2}\mathrm{Tr}\left[ \hat{\rho}_{%
\boldsymbol{\theta }}\left( \hat{L}_{\nu }^{(0)}\hat{L}_{\mu }^{(0)}+\hat{L}%
_{\mu }^{(0)}\hat{L}_{\nu }^{(0)}\right) \right] =F_{\mu \nu }^{(0)},
\label{smeq:Fmunv}
\end{equation}%
and thus we obtain the longitudinal block CFIM $\bm{\mathcal{F}}^{(0)}=\mathbf{F}^{(0)}$. Combining Eqs.\eqref{smeq:Faa}, %
\eqref{smeq:Fab}, \eqref{smeq:Fmunv}, and \eqref{smeq:Famu}, we obtain
\begin{equation}
\bm{\mathcal{F}}=%
\begin{pmatrix}
\bm{\mathcal{F}}^{(1)} & \mathbf{0} \\
\mathbf{0} & \bm{\mathcal{F}}^{(0)}%
\end{pmatrix}%
=%
\begin{pmatrix}
\mathbf{F}^{(1)} & \mathbf{0} \\
\mathbf{0} & \mathbf{F}^{(0)}%
\end{pmatrix}%
=\mathbf{F}.  \label{smeq:CFIMQFIM}
\end{equation}

\subsubsection{III. Extension to a three-sector block theorem}

\begin{figure}[hptb]
\centering
\includegraphics[width=0.52\columnwidth]{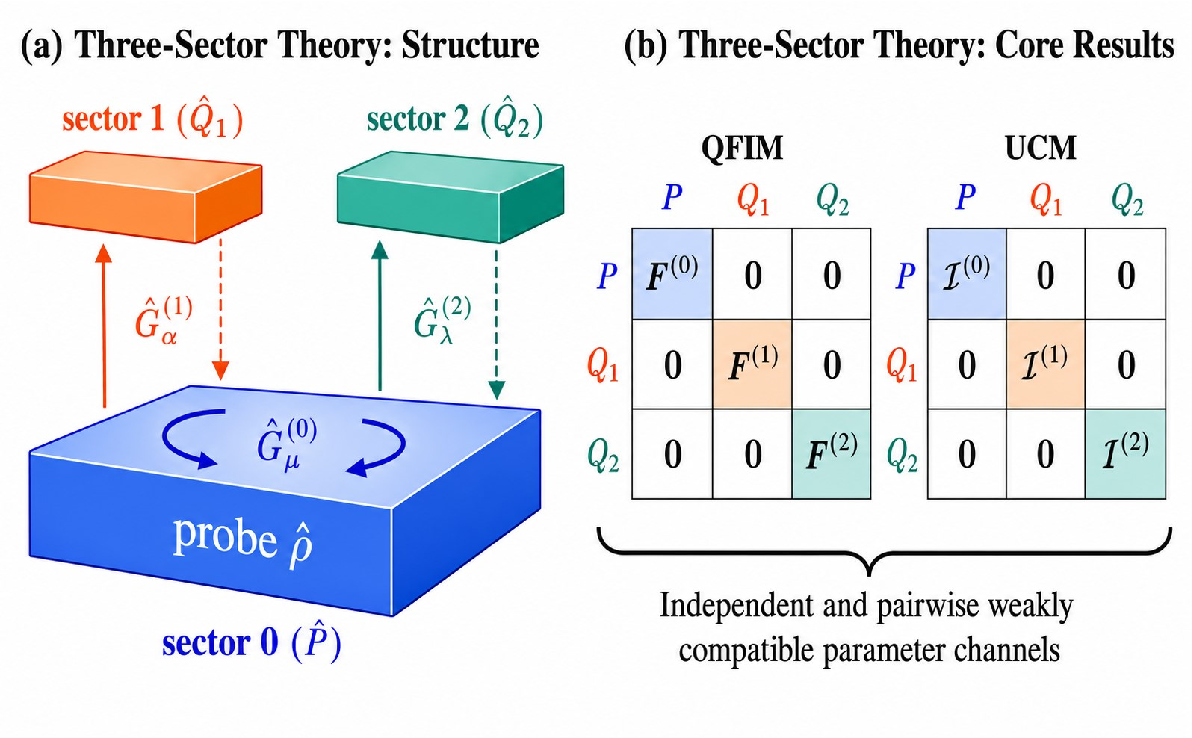}
\caption{ \textbf{Three-sector symmetry-projection mechanism.} (a) The probe
$\hat\protect\rho$ is supported in the reference sector associated with $%
\hat P$, while $\hat Q_1$ and $\hat Q_2$ denote two mutually orthogonal
complementary sectors. The longitudinal generators $\hat G_{\protect\mu%
}^{(0)}$ preserve the probe-supporting sector, whereas the two transverse
families $\hat G_{\protect\alpha}^{(1)}$ and $\hat G_{\protect\lambda}^{(2)}$
connect $\hat P$ independently to $\hat Q_1$ and $\hat Q_2$, respectively.
(b) The QFIM $\mathbf{F}$ and the UCM $\boldsymbol{\mathcal{I}}$ decompose
into the diagonal channel blocks $\mathbf{F}^{(a)}$ and $\boldsymbol{%
\mathcal{I}}^{(a)}$, $a=0,1,2$, while all cross-channel blocks vanish. The
three channels are therefore Fisher-orthogonal and weakly compatible across
different sectors, although compatibility within a multidimensional diagonal
block must be examined separately.}
\label{sm:three-sector}
\end{figure}

The sector-resolved mechanism developed above is not restricted to the
two-sector decomposition $\mathcal{H}=\mathcal{H}_{P}\oplus \mathcal{H}_{Q}$%
. It extends naturally to several mutually orthogonal complementary sectors,
where distinct intersector transitions can encode independently estimable
parameter channels. As illustrated in Fig.~\ref{sm:three-sector}, we
consider
\begin{equation}
\mathcal{H}=\mathcal{H}_{P}\oplus \mathcal{H}_{Q_{1}}\oplus \mathcal{H}%
_{Q_{2}},\qquad \hat{P}+\hat{Q}_{1}+\hat{Q}_{2}=\mathbb{I},
\label{eq:three-sector-decomposition}
\end{equation}%
where the projectors are pairwise orthogonal,
\begin{equation}
\hat{P}\hat{Q}_{a}=0,\qquad \hat{Q}_{1}\hat{Q}_{2}=0,\qquad \hat{P}^{2}=\hat{%
P},\qquad \hat{Q}_{a}^{2}=\hat{Q}_{a},\qquad (a=1,2).
\label{eq:three-sector-projectors}
\end{equation}%
The probe is fully supported in the reference sector,
\begin{equation}
\hat{\rho}=\hat{P}\hat{\rho}=\hat{\rho}\hat{P}=\hat{P}\hat{\rho}\hat{P}.
\label{eq:three-sector-probe}
\end{equation}%
Following the notation of the two-sector theorem, we introduce one
subspace-preserving family $\{\hat{G}_{\mu }^{(0)}\}$ and two
subspace-changing families $\{\hat{G}_{\alpha }^{(1)}\}$ and $\{\hat{G}%
_{\lambda }^{(2)}\}$. These generators are taken to be Hermitian. As
depicted in Fig.~\ref{sm:three-sector}, the indices are assigned according
to the subspace decomposition: $(\mu ,\nu ,...)$, $(\alpha ,\beta ,...)$,
and $(\lambda ,\kappa ,...)$ label the basis states of the $\hat{P}$, $\hat{Q%
}_{1}$, and $\hat{Q}_{2}$ subspaces, respectively, corresponding to the
sectors labeled by $a=0$, $1$, $2$.

The subspace-preserving property of $\hat{G}_{\mu }^{(0)}$ is expressed as
\begin{equation}
\lbrack \hat{G}_{\mu }^{(0)},\hat{P}]=0.  \label{eq:three-longitudinal}
\end{equation}%
For each subspace-changing family $a=1,2$, we impose the \emph{%
scalar-compression} condition
\begin{equation}
\hat{P}\hat{G}_{j}^{(a)}\hat{P}=g_{j}^{(a)}\hat{P},\qquad \left(
g_{j}^{(a)}\in \mathbb{R}\right) ,  \label{eq:three-scalar-compression}
\end{equation}%
and the subspace-resolved \emph{no-leakage} condition
\begin{equation}
\hat{Q}_{b}\hat{G}_{j}^{(a)}\hat{P}=\hat{P}\hat{G}_{j}^{(a)}\hat{Q}%
_{b}=0,\qquad a,b\in \{1,2\},\qquad (b\neq a).  \label{eq:three-no-leakage}
\end{equation}%
Here $j$ denotes the basis index within the corresponding subspace-changing
family. Unlike the two-sector case, where $\hat{Q}=\mathbb{I}-\hat{P}$ is
the unique complementary sector, Eq.~\eqref{eq:three-no-leakage} is a
genuine additional condition. It ensures that the $a$th subspace-changing
generator activates only $\mathcal{H}_{Q_{a}}$ and does not populate the
other complementary sector. Using the completeness relation $\hat{P}+\hat{Q}%
_{1}+\hat{Q}_{2}=\mathbb{I}$ and hence $\hat{Q}_{b}=\mathbb{I}-\hat{P}-\hat{Q%
}_{a}$, the no-leakage condition is equivalent to
\begin{equation}
(\mathbb{I}-\hat{P})\hat{G}_{j}^{(a)}\hat{P}=\hat{Q}_{a}\hat{G}_{j}^{(a)}%
\hat{P}.  \label{eq:three-no-leakage-equivalent}
\end{equation}%
A subspace-changing generator is nontrivial if it has a nonvanishing
coupling between the reference $\hat{P}$ subspace and the corresponding $%
\hat{Q}_{a}$ subspace:
\begin{equation}
\hat{Q}_{a}\hat{G}_{j}^{(a)}\hat{P}\neq 0.
\end{equation}%
By construction, the no-leakage condition ensures that the $\hat{Q}_{1}$ and
$\hat{Q}_{2}$ subspaces are not directly coupled. The scalar compression and
Eq.~\eqref{eq:three-sector-probe} imply
\begin{equation}
\left\langle \hat{G}_{j}^{(a)}\right\rangle =\mathrm{Tr}\left( \hat{\rho}%
\hat{P}\hat{G}_{j}^{(a)}\hat{P}\right) =g_{j}^{(a)},
\label{eq:three-expectation}
\end{equation}%
and hence the centered generators satisfy
\begin{equation}
\Delta \hat{G}_{j}^{(a)}\equiv \hat{G}_{j}^{(a)}-\left\langle \hat{G}%
_{j}^{(a)}\right\rangle ,\qquad \hat{P}\Delta \hat{G}_{j}^{(a)}\hat{P}=0.
\label{eq:three-centered-generators}
\end{equation}%
Each centered subspace-changing generator $\Delta \hat{G}_{j}^{(a)}$ has
nontrivial off-diagonal support between $\hat{P}$ and $\hat{Q}_{a}$
subspaces, with $\hat{P}\Delta \hat{G}_{j}^{(a)}\hat{Q}_{a}\neq 0$.
Specifically, using the no-leakage condition and the completeness relation,
we obtain%
\begin{eqnarray*}
\hat{P}\Delta \hat{G}_{j}^{(a)}\hat{Q}_{a} &=&\hat{P}\Delta \hat{G}%
_{j}^{(a)}(\mathbb{I}-\hat{P}-\hat{Q}_{b})=\hat{P}\Delta \hat{G}_{j}^{(a)}-%
\hat{P}\Delta \hat{G}_{j}^{(a)}\hat{P}-\hat{P}\Delta \hat{G}_{j}^{(a)}\hat{Q}%
_{b} \\
&=&\hat{P}\Delta \hat{G}_{j}^{(a)}-\hat{P}\left( \hat{G}_{j}^{(a)}-\left%
\langle \hat{G}_{j}^{(a)}\right\rangle \right) \hat{Q}_{b}=\hat{P}\Delta
\hat{G}_{j}^{(a)},
\end{eqnarray*}%
where we used $\hat{P}\Delta \hat{G}_{j}^{(a)}\hat{P}=0$, $\hat{P}\hat{G}%
_{j}^{(a)}\hat{Q}_{b}=0$, and $\hat{P}\hat{Q}_{b}=0$, arising from the
property of the centered generators Eq.~\eqref{eq:three-centered-generators}%
, the no-leakage condition Eq.~\eqref{eq:three-no-leakage}, and the mutual
orthogonality of the projection operators Eq.~%
\eqref{eq:three-sector-projectors}. Thus, we have
\begin{equation}
\hat{P}\Delta \hat{G}_{j}^{(a)}\hat{Q}_{a}=\hat{P}\Delta \hat{G}%
_{j}^{(a)},\qquad \hat{Q}_{a}\Delta \hat{G}_{j}^{(a)}\hat{P}=\Delta \hat{G}%
_{j}^{(a)}\hat{P}  \label{eq:three-active-support}
\end{equation}%
These relations show that the centered generators couple the reference
subspace only to the corresponding subspace, with no component leaking into
the other complementary subspace.

Under these assumptions, valid SLD representatives may be chosen as
\begin{equation}
\hat{L}_{\mu }^{(0)}=\hat{P}\hat{L}_{\mu }^{(0)}\hat{P}
\label{eq:three-longitudinal-sld}
\end{equation}%
for the longitudinal family, and
\begin{equation}
\hat{L}_{j}^{(a)}=-2i\left[ \Delta \hat{G}_{j}^{(a)},\hat{P}\right]
=2i\left( \hat{P}\Delta \hat{G}_{j}^{(a)}\hat{Q}_{a}-\hat{Q}_{a}\Delta \hat{G%
}_{j}^{(a)}\hat{P}\right) ,\qquad (a=1,2),  \label{eq:three-transverse-sld}
\end{equation}%
for the two transverse families. The chosen SLDs have the support structure
\begin{equation}
\hat{\boldsymbol{L}}^{(0)}:\mathcal{H}_{P}\longrightarrow \mathcal{H}%
_{P},\qquad \hat{\boldsymbol{L}}^{(1)}:\mathcal{H}_{P}\leftrightarrow
\mathcal{H}_{Q_{1}},\qquad \hat{\boldsymbol{L}}^{(2)}:\mathcal{H}%
_{P}\leftrightarrow \mathcal{H}_{Q_{2}}.
\label{eq:three-sld-support-actions}
\end{equation}%
Therefore, for distinct channels $a\neq b$,
\begin{equation}
\hat{P}\hat{L}_{j}^{(a)}\hat{L}_{k}^{(b)}\hat{P}=0.
\label{eq:three-sld-cross-product}
\end{equation}%
Since $\hat{\rho}=\hat{P}\hat{\rho}\hat{P}$, it follows that
\begin{equation}
\mathrm{Tr}\left( \hat{\rho}\hat{L}_{j}^{(a)}\hat{L}_{k}^{(b)}\right)
=0,\qquad (a\neq b).  \label{eq:three-cross-trace}
\end{equation}%
Consequently, both the symmetrized and antisymmetrized cross-channel SLD
products vanish:
\begin{align}
F_{jk}^{(a,b)}& =\frac{1}{2}\mathrm{Tr}\left[ \hat{\rho}\left\{ \hat{L}%
_{j}^{(a)},\hat{L}_{k}^{(b)}\right\} \right] =0,  \label{eq:three-cross-qfim}
\\
\mathcal{I}_{jk}^{(a,b)}& =\frac{1}{2i}\mathrm{Tr}\left[ \hat{\rho}\left[
\hat{L}_{j}^{(a)},\hat{L}_{k}^{(b)}\right] \right] =0.
\label{eq:three-cross-ucm}
\end{align}%
Here $j$, $k$ label the generators within the corresponding SLD families $a$
and $b$. As illustrated in Fig.~\ref{sm:three-sector}(b), the QFIM and UCM
then have the direct-sum forms
\begin{equation}
\mathbf{F}=\mathbf{F}^{(0)}\oplus \mathbf{F}^{(1)}\oplus \mathbf{F}%
^{(2)},\qquad \boldsymbol{\mathcal{I}}=\boldsymbol{\mathcal{I}}^{(0)}\oplus
\boldsymbol{\mathcal{I}}^{(1)}\oplus \boldsymbol{\mathcal{I}}^{(2)},
\label{eq:three-block-matrices}
\end{equation}%
or equivalently,
\begin{equation}
\mathbf{F}=%
\begin{pmatrix}
\mathbf{F}^{(0)} & \mathbf{0} & \mathbf{0} \\
\mathbf{0} & \mathbf{F}^{(1)} & \mathbf{0} \\
\mathbf{0} & \mathbf{0} & \mathbf{F}^{(2)}%
\end{pmatrix}%
,\qquad \boldsymbol{\mathcal{I}}=%
\begin{pmatrix}
\boldsymbol{\mathcal{I}}^{(0)} & \mathbf{0} & \mathbf{0} \\
\mathbf{0} & \boldsymbol{\mathcal{I}}^{(1)} & \mathbf{0} \\
\mathbf{0} & \mathbf{0} & \boldsymbol{\mathcal{I}}^{(2)}%
\end{pmatrix}%
.  \label{eq:three-explicit-block-matrices}
\end{equation}%
Thus, all cross-channel blocks vanish:
\begin{equation}
\mathbf{F}^{(a,b)}=0,\qquad \boldsymbol{\mathcal{I}}^{(a,b)}=0,\qquad (a\neq
b).  \label{eq:three-cross-blocks}
\end{equation}
This result also admits a simple tangent-space interpretation: the tangent
directions associated with different subspace families are mutually
orthogonal. This orthogonality is the geometric origin of the vanishing real
and imaginary parts of the cross-family SLD pairings.

To express the QFIM contribution from the subspace-changing families in
terms of generator covariances, we first simplify the product of two SLDs
for a fixed family. Using Eq.~\eqref{eq:three-transverse-sld}, the support
structure gives%
\begin{equation*}
\hat{P}\hat{L}_{j}^{(a)}\hat{L}_{k}^{(a)}\hat{P}=-4\hat{P}\left( -\hat{P}%
\Delta \hat{G}_{j}^{(a)}\hat{Q}_{a}\Delta \hat{G}_{k}^{(a)}\hat{P}\right)
\hat{P}=4\hat{P}\Delta \hat{G}_{j}^{(a)}\left( \hat{Q}_{a}\Delta \hat{G}%
_{k}^{(a)}\hat{P}\right) =4\hat{P}\Delta \hat{G}_{j}^{(a)}\Delta \hat{G}%
_{k}^{(a)}\hat{P},
\end{equation*}%
where we used the support structure of the transverse SLDs, the mutual
orthogonality and idempotency of the projection operators, and the
centered-generator property Eq.~\eqref{eq:three-active-support}, i.e., $\hat{%
Q}_{a}\Delta \hat{G}_{k}^{(a)}\hat{P}=\Delta \hat{G}_{k}^{(a)}\hat{P}$.
Consequently, we have%
\begin{eqnarray}
\mathrm{Tr}\left( \hat{\rho}\hat{L}_{j}^{(a)}\hat{L}_{k}^{(a)}\right) &=&%
\mathrm{Tr}\left( \hat{P}\hat{\rho}\hat{P}\hat{L}_{j}^{(a)}\hat{L}%
_{k}^{(a)}\right) =\mathrm{Tr}\left( \hat{\rho}\underbrace{\hat{P}\hat{L}%
_{j}^{(a)}\hat{L}_{k}^{(a)}\hat{P}}\right) =4\mathrm{Tr}\left( \hat{\rho}%
\underbrace{\hat{P}\Delta \hat{G}_{j}^{(a)}\Delta \hat{G}_{k}^{(a)}\hat{P}}%
\right)  \notag \\
&=&4\mathrm{Tr}\left( \hat{P}\hat{\rho}\hat{P}\Delta \hat{G}_{j}^{(a)}\Delta
\hat{G}_{k}^{(a)}\right) =4\mathrm{Tr}\left( \hat{\rho}\Delta \hat{G}%
_{j}^{(a)}\Delta \hat{G}_{k}^{(a)}\right) =4\left\langle \Delta \hat{G}%
_{j}^{(a)}\Delta \hat{G}_{k}^{(a)}\right\rangle .
\end{eqnarray}%
Taking the symmetrized combination with respect to $j$ and $k$, we obtain
the QFIM for the subspace-changing families:
\begin{equation}
F_{jk}^{(a)}=\frac{1}{2}\mathrm{Tr}\left( \hat{\rho}\left\{ \hat{L}%
_{j}^{(a)},\hat{L}_{k}^{(a)}\right\} \right) =4\mathrm{Cov}\left( \hat{G}%
_{j}^{(a)},\hat{G}_{k}^{(a)}\right) ,\qquad (a=1,2),
\label{eq:three-covariance-proof}
\end{equation}%
where where $\mathrm{Cov}(\hat{A},\hat{B})\equiv \frac{1}{2}\langle \{\Delta
\hat{A},\Delta \hat{B}\}\rangle $, with $\Delta \hat{O}\equiv \hat{O}%
-\langle \hat{O}\rangle $.

The three-sector framework extends naturally to multiple orthogonal sectors,
allowing several independently addressable sensing channels to be treated
within the same symmetry-projected geometry. The resulting QFIM and Uhlmann
curvature matrix retain their block structure across distinct sectors, while
compatibility within each multidimensional block can be analyzed separately.
This generalization opens a route to multiparameter sensing in systems with
multiple excitation or transition channels, such as higher-spin condensates,
multimode trapped-ion platforms, multimode cavity-QED systems, and
superconducting circuits.

\subsubsection{IV. Unified Symmetry-Projected Sensing Geometry and Key Consequences}

Taken together, the two- and three-sector constructions unveil a unified symmetry-protected geometry for multiparameter quantum estimation: symmetry projection induces a common block-diagonal structure on both the quantum Fisher information matrix (QFIM) and the Uhlmann curvature matrix (UCM), renders distinct subspace-changing channels Fisher-orthogonal and weakly compatible, and simplifies the associated QFIM blocks to symmetrized generator covariances. This algebraic structure persists across multiple independent tangent channels, providing a versatile blueprint for general multi-sector sensing architectures. The main operational consequences are summarized as follows:

\begin{enumerate}
\item \textbf{Block diagonalization and elimination of information cross-talk.} At the reference parameter point, distinct parameter sectors are locally decoupled at the QFIM level, $\mathbf{F}^{(a,b)}=\mathbf{0}$ for $a\neq b$. The resulting block-diagonal QFIM guarantees complete freedom from quantum information cross-talk between different sensing channels.

\item \textbf{Direct variance-based sensitivity evaluation.} When a subspace-changing generator satisfies scalar compression on the probe support, the associated QFIM block reduces exactly to four times the symmetrized covariance matrix,
\begin{equation*}
F_{\alpha\beta}^{(a)} = 4\,\mathrm{Cov} \!\left( \hat G_{\alpha}^{(a)}, \hat G_{\beta}^{(a)} \right).
\end{equation*}
Consequently, quantum sensitivities can be directly evaluated from two-point fluctuation moments, bypassing the need for full quantum state tomography or spectral decomposition of mixed probe states.

\item \textbf{Hierarchy of compatibility and multi-channel decoupling.} The framework explicitly establishes a three-tier compatibility hierarchy:
(i) Fisher orthogonality ($\mathbf{F}^{(a,b)}=\mathbf{0}$ for $a\neq b$);
(ii) cross-sector weak compatibility ($\boldsymbol{\mathcal{I}}^{(a,b)}=\mathbf{0}$ for $a\neq b$); and
(iii) full weak compatibility ($\boldsymbol{\mathcal{I}}=\mathbf{0}$ across all parameters).
Symmetry projection strictly enforces the first two properties. Full weak compatibility automatically holds if at most one parameter is chosen from each channel, as every single-parameter diagonal curvature block identically vanishes by antisymmetry.

\item \textbf{Attainability of precision bounds and platform universality.} Cross-sector weak compatibility ensures that coupling between distinct sensing channels introduces no trade-off penalty. If individual diagonal blocks are also weakly compatible ($\boldsymbol{\mathcal{I}}=\mathbf{0}$), the total Uhlmann curvature vanishes, causing the Holevo bound to coincide with the SLD Cram\'er-Rao bound, which is asymptotically achievable via collective measurements in the multi-copy limit. Crucially, this compatibility requires neither a joint SLD eigenbasis nor global commutativity of the encoding generators. Dependent solely on the probe support and symmetry-resolved generator algebra, this framework is independent of Hilbert-space dimension and applies directly to spin ensembles, bosonic platforms, and open quantum many-body systems.

\end{enumerate}

\subsection*{Application to parity-protected collective SU(2) systems}

We consider a particle-number-conserving collective $\mathrm{SU}(2)$ system,
where the phase generators $\hat{\mathbf{G}}=\hat{\mathbf{J}}=(\hat{J}_{x},%
\hat{J}_{y},\hat{J}_{z})$. Choosing the standard single-particle basis $%
\{|\!\uparrow \rangle ,|\!\downarrow \rangle \}$, we define the spin
operators $\hat{J}_{z}=\frac{1}{2}(\hat{N}_{\uparrow }-\hat{N}_{\downarrow
}) $ and $\hat{J}_{\pm }=\hat{J}_{x}\pm i\hat{J}_{y}$. For an ensemble of $N$
spin-$1/2$ particles (equivalent to a singe large spin with total spin $%
J=N/2 $), we have
\begin{equation}
\hat{N}_{\uparrow }=\sum_{n=1}^{N}|\!\uparrow \rangle _{n}\,{}_{n}\langle
\uparrow \!|,\qquad \hat{N}_{\downarrow }=\sum_{n=1}^{N}|\!\downarrow
\rangle _{n}\,{}_{n}\langle \downarrow \!|,
\end{equation}
and $\hat{J}_{+}=(\hat{J}_{-})^{\dag }=\sum_{n=1}^{N}|\!\uparrow \rangle
_{n}\,{}_{n}\langle \downarrow \!|$. For a two-mode boson system, we adopt
Schwinger's representation to define $\hat{N}_{\uparrow }=\hat{a}_{\uparrow
}^{\dagger }\hat{a}_{\uparrow }$, $\hat{N}_{\downarrow}=\hat{a}_{\downarrow
}^{\dagger }\hat{a}_{\downarrow }$, and $\hat{J}_{+}=(\hat{J}_{-})^{\dag }=%
\hat{a}_{\uparrow }^{\dagger }\hat{a}_{\downarrow }$. With fixed total
particle number $N$, we have $\hat{N}_{\uparrow }+\hat{N}_{\downarrow}=N=2J$%
. Thus, the parity operator with respect to the $z$-axis is defined by
\begin{equation}
\hat{\Pi}=(-1)^{\hat{N}_{\uparrow }}=\exp \!\left[ i\pi \left( J+\hat{J}%
_{z}\right) \right] .  \label{smeq:parity}
\end{equation}%
where $\hat{\Pi}^{2}=\mathbb{I}$. Note that the spin operators satisfy $[%
\hat{J}_{z},\hat{\Pi}]=0$, while $\{\hat{J}_{x,y},\hat{\Pi}\}=0$, due to $%
e^{\pm i\pi \hat{J}_{z}}\hat{J}_{x,y}e^{\mp i\pi \hat{J}_{z}}=-\hat{J}_{x,y}$%
. Therefore, $(\hat{J}_{x},\hat{J}_{y})$ flip the parity, whereas $\hat{J}%
_{z}$ preserves the parity.

\subsubsection*{I. Spectral decomposition of the QFIM}

The QFIM elements can be expressed explicitly via spectral decomposition of
arbitrary spin state $\hat{\rho}$ (see e.g., Refs.~\cite%
{Liu2020_supp,Hyllus_supp})
\begin{eqnarray}
F_{\alpha \beta } &=&\sum_{k,l}\frac{2(\lambda _{k}-\lambda _{l})^{2}}{%
\lambda _{k}+\lambda _{l}}\langle k|\hat{J}_{\alpha }|l\rangle \langle l|%
\hat{J}_{\beta }|k\rangle \\
&=&2\sum_{k,l}(\lambda _{k}+\lambda _{l})\langle k|\hat{J}_{\alpha
}|l\rangle \langle l|\hat{J}_{\beta }|k\rangle -\sum_{k,l}\frac{8\lambda
_{k}\lambda _{l}}{\lambda _{k}+\lambda _{l}}\langle k|\hat{J}_{\alpha
}|l\rangle \langle l|\hat{J}_{\beta }|k\rangle  \notag \\
&=&2\sum_{k}\lambda _{k}\langle k|(\hat{J}_{\alpha }\hat{J}_{\beta }+\hat{J}%
_{\beta }\hat{J}_{\alpha })|k\rangle -\sum_{k,l}\frac{8\lambda _{k}\lambda
_{l}}{\lambda _{k}+\lambda _{l}}\langle k|\hat{J}_{\alpha }|l\rangle \langle
l|\hat{J}_{\beta }|k\rangle  \notag \\
&=&2\left\langle \left\{ \hat{J}_{\alpha },\hat{J}_{\beta }\right\}
\right\rangle -\sum_{k,l}\frac{8\lambda _{k}\lambda _{l}}{\lambda
_{k}+\lambda _{l}}\langle k|\hat{J}_{\alpha }|l\rangle \langle l|\hat{J}%
_{\beta }|k\rangle ,  \label{smeq:QFIM}
\end{eqnarray}%
where $\alpha ,\beta =x,y,z$, and the expetation value $\langle \hat{O}%
\rangle =\mathrm{Tr}(\hat{\rho}\hat{O})=\sum_{k}\lambda _{k}\langle k|\hat{O}%
|k\rangle $, dependent on the spectral decomposition of the probe state%
\begin{equation}
\hat{\rho}=\sum_{k}\lambda _{k}|k\rangle \langle k|,\qquad \sum_{k}\lambda
_{k}=1,  \label{smeq:spectral decomposition}
\end{equation}%
with eigenstates $\{|k\rangle \}$ and eigenvalues $\lambda _{k}>0$. The
double-sum correction in Eq.~\eqref{smeq:QFIM} vanishes if and only if $%
\lambda _{k}\lambda _{l}=0$ for all pairs of eigenstates satisfying $\langle
k|\hat{J}_{\alpha }|l\rangle \neq 0$.

For a probe state fully supported within a definite parity sector, $\hat{\rho%
}=\hat{P}\hat{\rho}\hat{P}$, the corresponding projection operator is $\hat{P%
}=(\mathbb{I}\pm \hat{\Pi})/2$, where the $+$ and $-$ signs denote the even-
and odd-parity sectors, respectively. The spectral decomposition of $\hat{%
\rho}$ in the $\hat{P}$ subspace takes the form of Eq.~%
\eqref{smeq:spectral
decomposition}, while in its orthogonal $\hat{Q}$ subspace, all eigenstates $%
\{|l\rangle \}$ have vanishing eigenvalues $\lambda _{l}=0$. Consequently,
the double-sum correction in Eq.~\eqref{smeq:QFIM} vanishes whenever at
least one index corresponds to the transverse generator $\hat{J}_{x,y}$.
This holds not only to the entire transverse block $(F_{xx},F_{xy},F_{yy})$,
but also to the cross-sector elements $(F_{xz},F_{yz})$. In contrast, the
parity-preserving longitudinal generator $\hat{J}_{z}$ retains finite
intra-sector matrix elements, yielding $F_{zz}\leq 4\mathrm{Var}(\hat{J}%
_{z}) $, which degrades for mixed states and requires the full mixed-state
QFIM evaluation.

The first term of Eq.~\eqref{smeq:QFIM} gives
\begin{equation}
F_{\alpha \beta }=2\langle \{\hat{J}_{\alpha },\hat{J}_{\beta }\}\rangle
,\qquad (\alpha ,\beta) \neq (z,z),  \label{smeq:transverse-block}
\end{equation}%
where the cross-sector terms
\begin{eqnarray}
F_{yz}&=&2\langle \{\hat{J}_{y },\hat{J}_{z}\}\rangle=2\mathrm{Im}\langle
\hat{J}_{+}(2\hat{J}_{z}+1)\rangle =0,  \label{smeq:Fyz} \\
F_{xz}&=&2\langle \{\hat{J}_{x},\hat{J}_{z}\}\rangle=2\mathrm{Re}\langle
\hat{J}_{+}(2\hat{J}_{z}+1)\rangle =0.  \label{smeq:Fxz}
\end{eqnarray}%
Here, we use the identity $\langle \hat{J}_{+}(2\hat{J}_{z}+1)\rangle =0$,
which stems from the parity-symmetry selection rule. Therefore, we obtain
the block-diagonal QFIM,
\begin{equation}
\mathbf{F}=%
\begin{pmatrix}
F_{xx} & F_{xy} & 0 \\
F_{xy} & F_{yy} & 0 \\
0 & 0 & F_{zz}%
\end{pmatrix}%
.  \label{smeq:block-diagonal}
\end{equation}%
For the transverse QFIM components, the parity symmetry enforces $\langle
\hat{J}_{+}\rangle =0$ and hence $\langle \hat{J}_{x}\rangle =\langle \hat{J}%
_{y}\rangle =0$. Equation~\eqref{smeq:transverse-block} therefore recovers
the pure-state relation~\cite{Hyllus_supp}: $F_{\alpha \beta }=4\mathrm{Cov}(%
\hat{J}_{\alpha },\hat{J}_{\beta })$, where $\mathrm{Cov}(\hat{A},\hat{B}%
)\equiv \frac{1}{2}\langle \hat{A}\hat{B}+\hat{B}\hat{A}\rangle -\langle
\hat{A}\rangle \langle \hat{B}\rangle $ denotes the symmetrized covariance
between two operators.

We have shown that the block-diagonal structure of the QFIM -- including its
transverse entries -- stems directly from parity symmetry, holding for any
mixed state satisfying $\hat{\rho}=\hat{P}\hat{\rho}\hat{P}$. By contrast,
the weak compatibility condition depends on the explicit form of the SLDs
and warrants a separate proof (see below).

\subsubsection*{II. Spin fluctuations of SU(2) collective spin}

Before addressing the weak compatibility condition, it is instructive to
highlight the physical significance of these transverse entries. The
relationship between the transverse QFIM components and the spin
fluctuations establishes an explicit operational link to spin-squeezing
protocols~\cite{Kitagawa_supp}, where the spin fluctuations normal to the
mean spin are of central interest in single-parameter quantum metrology~\cite%
{Wineland1_supp,Wineland1994_supp}.

We characterize the transverse spin squeezing and phase sensing properties
in the main text using the transverse spin components, their fluctuations,
and the corresponding symmetrized covariance matrix. Here, we provide
complementary analyses based on the conventional spin-squeezing formalism~%
\cite{Kitagawa_supp,JinNJP09_supp} and the single-parameter QFI optimization~%
\cite{Hyllus_supp}, with the optimal squeezing angle and the optimal sensing
direction discussed separately. These approaches yield results fully
consistent with those presented in the main text.

\textit{Spin squeezing.---}For the parity-protected state $\hat{\rho}$, the
parity-symmetry selection rule $\langle \hat{J}_{+}\rangle =0$ ensures the
mean spin vector $\langle \hat{\mathbf{J}}\rangle =(0,0,\langle \hat{J}%
_{z}\rangle )$, pointing strictly along the $z$ axis as $\langle \hat{J}%
_{z}\rangle \neq 0$. Consequently, a generic spin component normal to the
mean spin is defined by $\hat{J}_{\eta }=\hat{\mathbf{J}}\cdot {\mathbf{n}}%
_{\bot }$, where ${\mathbf{n}}_{\bot }=(\cos \eta ,\sin \eta ,0)$. Since $%
\langle \hat{J}_{\eta }\rangle =0$, its variance is given by
\begin{equation}
\langle \hat{J}_{\eta }^{2}\rangle =\frac{1}{2}\left( \langle \hat{C}\rangle
+\langle \hat{A}\rangle \cos (2\eta )+\langle \hat{B}\rangle \sin (2\eta
)\right) ,
\end{equation}%
where we introduce the correlation operators
\begin{equation}
\hat{A}=\hat{J}_{x}^{2}-\hat{J}_{y}^{2},\qquad \hat{B}=\hat{J}_{x}\hat{J}%
_{y}+\hat{J}_{y}\hat{J}_{x},\qquad \hat{C}=\hat{J}_{x}^{2}+\hat{J}_{y}^{2}.
\label{smeq:ABC}
\end{equation}%
Maximizing the variance with respect to $\eta$, we obtain the optimal
squeezing angle $\eta _{\mathrm{op}}$, satisfying $\partial \langle \hat{J}%
_{\eta }^{2}\rangle /\partial \eta |_{\eta _{\mathrm{op}}}=0$ and hence $%
\tan (2\eta _{\mathrm{op}})=\langle \hat{B}\rangle /\langle \hat{A}\rangle $%
. Therefore, the squeezed and anti-squeezed variances relative to the
standard quantum limit ($N/4$) are given by~\cite%
{Kitagawa_supp,JinNJP09_supp},
\begin{equation}
V_{\pm }=\frac{1}{2}\left( \langle \hat{C}\rangle \pm \sqrt{\langle \hat{A}%
\rangle ^{2}+\langle \hat{B}\rangle ^{2}}\right) ,  \label{smeq:V+-}
\end{equation}%
with the associated squeezing and anti-squeezing directions
\begin{equation}
\mathbf{n}_{-}=(\cos \eta _{\mathrm{op}},\sin \eta _{\mathrm{op}},0),\qquad
\mathbf{n}_{+}=(-\sin \eta _{\mathrm{op}},\cos \eta _{\mathrm{op}},0).
\label{smeq:n+-}
\end{equation}

\textit{Single-parameter QFI optimization.---}Here, the parity symmetry of
the probe enforces a transverse--longitudinal block structure in the QFIM.
Its principal eigenvectors therefore lie within the corresponding generator
sectors. This reduction avoids an explicit spectral analysis of the probe in
the full Hilbert space. To illustrate this reduction, we consider optimal
single-parameter phase sensing with a fixed probe state $\hat{\rho}$,
encoded as
\begin{equation}
\hat{\rho}_{\theta }=e^{-i\theta \hat{G}}\hat{\rho}e^{i\theta \hat{G}%
},\qquad \hat{G}=\hat{\mathbf{J}}\cdot \mathbf{n}_{s}, \qquad (\mathbf{n}_{s}%
\mathbf{n}_{s}^{T}=1).  \label{smeq:spUnitary}
\end{equation}
Since $F_Q=\mathbf{n}_{s}\mathbf{F}\mathbf{n}_{s}^{T}$, optimization over
the three-dimensional sensing axis reduces to diagonalizing the QFIM~\cite%
{Reilly_supp}. Its largest eigenvalue gives the maximal QFI, and the
associated eigenvector determines the optimal sensing direction~\cite%
{Hyllus_supp}. By diagonalizing Eq.~\eqref{smeq:block-diagonal}, we obtain
the maximal QFI and its associated optimal sensing direction
\begin{equation}
F_{Q,\max}=\max \{4V_{+},F_{zz}\},\qquad \mathbf{n}_{s,\max }\in \{{\mathbf{n%
}}_{+},\hat{z}\},  \label{smeq:qfi}
\end{equation}%
as Eq.~(16) in main text. In the derivation, we have expressed the
transverse QFIM components in terms of correlators:
\begin{equation}
F_{xx}=4\langle \hat{J}_{x}^{2}\rangle =2\langle \hat{C}+\hat{A}\rangle
,\quad F_{yy}=4\langle \hat{J}_{y}^{2}\rangle =2\langle \hat{C}-\hat{A}%
\rangle ,\quad F_{xy}=2\langle \{\hat{J}_{x},\hat{J}_{y}\}\rangle =2\langle
\hat{B}\rangle ,  \label{smeq:FABC}
\end{equation}%
where $\langle \hat{A}\rangle =\mathrm{Re}\langle \hat{J}_{+}^{2}\rangle $, $%
\langle \hat{B}\rangle =\mathrm{Im}\langle \hat{J}_{+}^{2}\rangle $, and $%
\langle \hat{C}\rangle =J(J+1)-\langle \hat{J}_{z}^{2}\rangle $. The above
results hold for any mixed state satisfying $\hat{\rho}=\hat{P}\hat{\rho}%
\hat{P}$. In single-parameter phase estimation, our analytical results
identify a competition between $F_{zz}$ and $4V_{+}$, dictating whether the
optimal sensing direction $\mathbf{n}_{s, \max }$ aligns with the $z$-axis
or the anti-squeezing direction $\mathbf{n}_{+}$. This is a direct
consequence of the parity symmetry, which eliminates all
transverse-longitudinal cross terms, reducing the full three-axis
optimization of the QFI to an exact comparison between $4V_{+}$ and $F_{zz}$.

\subsubsection*{III. Weak Compatibility of parity-protected SU(2) Systems}

The multi-parameter quantum channel encodes the parameters $\boldsymbol{%
\theta }=(\theta _{x},\theta _{y})$ via the unitary transformation
\begin{equation}
\hat{U}(\boldsymbol{\theta })=e^{-i(\theta _{x}\hat{J}_{x}+\theta _{y}\hat{J}%
_{y})}.
\end{equation}%
Defining $\hat{A}(\boldsymbol{\theta })=-i(\theta _{x}\hat{J}_{x}+\theta _{y}%
\hat{J}_{y})$ such that $\hat{U}(\boldsymbol{\theta })=e^{\hat{A}(%
\boldsymbol{\theta })}$, we evaluate the local response of the
parameter-dependent density matrix $\hat{\rho}_{\boldsymbol{\theta }}=\hat{U}%
(\boldsymbol{\theta })\hat{\rho}\hat{U}^{\dagger }(\boldsymbol{\theta })$ at
the reference point $\boldsymbol{\theta }=0$. We employ the Wilcox identity
for the derivative of an exponential operator~\cite{Liu2020_supp},
\begin{equation}
\frac{\partial \hat{U}(\boldsymbol{\theta }) }{\partial \theta _{\alpha }}=%
\frac{\partial }{\partial \theta _{\alpha }}e^{\hat{A}(\boldsymbol{\theta }%
)}=\int_{0}^{1}e^{s\hat{A}}\frac{\partial \hat{A}}{\partial \theta _{\alpha }%
}e^{(1-s)\hat{A}}ds,  \label{smeq:wilcox}
\end{equation}%
where $\partial \hat{A}/\partial \theta _{\alpha }=-i\hat{J}_{\alpha }$,
independent of $\boldsymbol{\theta }$. Evaluating Eq.~\eqref{smeq:wilcox} at
the reference point $\boldsymbol{\theta }=0$ collapses the exponential
factors to the identity, $e^{s\hat{A}(0)}=\mathbb{I}$, since $\hat{A}(0)=0$.
Therefore, the derivative at $\boldsymbol{\theta }=0$ simplifies to the
standard commutator form
\begin{equation}
\partial _{\alpha }\hat{\rho}=\left. \frac{\partial \hat{\rho}_{\boldsymbol{%
\theta }}}{\partial \theta _{\alpha }}\right\vert _{\boldsymbol{\theta }%
=0}=-i[\hat{J}_{\alpha },\hat{\rho}],
\end{equation}%
where $\partial _{\alpha }=\partial /\partial \theta _{\alpha }$. The SLD
operators $\hat{L}_{x,y}$ for parameters $\theta _{x,y}$ are implicitly
defined via the Lyapunov equation~\cite{Liu2020_supp}:
\begin{equation}
\partial _{\alpha }\hat{\rho}=\frac{1}{2}\{\hat{L}_{\alpha },\hat{\rho}\}.
\label{smeq:Lyapunov}
\end{equation}%
In the eigenbasis of $\hat{\rho}$, the matrix elements of $\hat{L}_{\alpha }$
are explicitly given by~\cite{Liu2020_supp}:
\begin{equation}
\langle \psi _{m}|\hat{L}_{\alpha }|\psi _{n}\rangle =\frac{2i(\lambda
_{m}-\lambda _{n})}{\lambda _{m}+\lambda _{n}}\langle \psi _{m}|\hat{J}%
_{\alpha }|\psi _{n}\rangle ,
\end{equation}%
where $|\psi _{m}\rangle ,|\psi _{n}\rangle $ are eigenstates of $\hat{\rho}$%
, with eigenvalues $\lambda _{m}+\lambda _{n}\neq 0$.

A fundamental distinction separates multi-parameter quantum estimation from
its single-parameter counterpart. When simultaneously estimating the
parameters, the ultimate precision limit is fundamentally constrained by
quantum incompatibility. In quantum information theory, this joint
estimation threshold is rigorously governed by the Holevo bound, where the
non-commuting nature introduces a strict quantum penalty~\cite%
{Liu2020_supp,Ragy2016_supp}. This incompatibility penalty vanishes under
the weak compatibility condition, which is quantified by the Uhlmann
curvature~\cite{Liu2020_supp}:
\begin{equation}
\mathcal{I}_{xy}\equiv\frac{1}{2i}\mathrm{Tr}\left( \hat{\rho}[\hat{L}_{x},%
\hat{L}_{y}]\right).  \label{smeq:weak_compat}
\end{equation}%
When $\mathcal{I}_{xy}=0$, the joint estimation precision can asymptotically
saturate the standard SLD Cram\'{e}r--Rao bound simultaneously for both
parameters, completely eliminating any quantum trade-off.

The spectral decomposition of $\hat{\rho}$ in the $\hat{P}$ subspace takes
the form of Eq.~\eqref{smeq:QFIM}. To clarify, we express $\hat{\rho}$ in
its orthogonal $\hat{Q}$ subspace, using different notations of eigenstates $%
\{|\widetilde{l}\rangle \}$ and their eigenvalues $\lambda _{\widetilde{l}}$
($=0$). The matrix elements of $\hat{L}_{\alpha }$ can be expressed as
\begin{eqnarray}
\langle k|\hat{L}_{\alpha }|\widetilde{l}\rangle &=&\frac{2i(\lambda _{k}-0)%
}{\lambda _{k}+0}\langle k|\hat{J}_{\alpha }|\widetilde{l}\rangle =2i\langle
k|\hat{J}_{\alpha }|\widetilde{l}\rangle , \\
\langle \widetilde{l}|\hat{L}_{\alpha }|k\rangle &=&\frac{2i(0-\lambda _{k})%
}{0+\lambda _{k}}\langle \widetilde{l}|\hat{J}_{\alpha }|k\rangle
=-2i\langle \widetilde{l}|\hat{J}_{\alpha }|k\rangle .
\end{eqnarray}%
Reconstructing the full operator from these distinct subspace blocks leads
to a universal, state-independent expression
\begin{equation}
\hat{L}_{\alpha }=2i\left( \hat{P}\hat{J}_{\alpha }\hat{Q}-\hat{Q}\hat{J}%
_{\alpha }\hat{P}\right) =-2i\left[ \hat{J}_{\alpha },\hat{P}\right] ,
\qquad(\alpha =x,y),  \label{Lalpha}
\end{equation}%
coincident with Eq.~\eqref{smeq:SLD1} and Eq.~(6) in main text. Considering
the even-parity state with $\hat{P}=(\mathbb{I}+\hat{\Pi})/2$, we have $\hat{%
\Pi}\hat{J}_{\alpha }=-\hat{J}_{\alpha }\hat{\Pi}$ (i.e., $\hat{\Pi}\hat{J}%
_{\alpha }\hat{\Pi}=-\hat{J}_{\alpha }$) and thus $[ \hat{J}_{\alpha },\hat{%
\Pi}] =2\hat{J}_{\alpha }\hat{\Pi}$, which in turn gives
\begin{equation}
\hat{L}_{\alpha }=-i\left[ \hat{J}_{\alpha },\hat{\Pi}\right] =-2i\hat{J}%
_{\alpha }\hat{\Pi}, \qquad(\alpha =x,y),  \label{LPi}
\end{equation}%
and%
\begin{align}
\lbrack \hat{L}_{x},\hat{L}_{y}]& =-4[\hat{J}_{x}\hat{\Pi},\hat{J}_{y}\hat{%
\Pi}]=-4\left( \hat{J}_{x}\hat{\Pi}\hat{J}_{y}\hat{\Pi}-\hat{J}_{y}\hat{\Pi}%
\hat{J}_{x}\hat{\Pi}\right)  \notag \\
& =-4\left( -\hat{J}_{x}\hat{J}_{y}+\hat{J}_{y}\hat{J}_{x}\right) =4[\hat{J}%
_{x},\hat{J}_{y}]=4i\hat{J}_{z}.  \label{LxLy}
\end{align}%
This holds for the odd-parity state with $\hat{P}=(\mathbb{I}-\hat{\Pi})/2$.
Substituting the two generators into Eq.~\eqref{smeq:weak_compat} gives
\begin{equation}
\mathcal{I}_{xy}=2\langle\hat J_z\rangle .  \label{Ixy_Jz}
\end{equation}
Thus, the transverse parameter pair $(\theta_x,\theta_y)$ is weakly
compatible if and only if the probe has vanishing magnetization along $z$.

For the even-parity steady state of the dissipative OAT model,
\begin{equation}
\hat{\rho}_{\mathrm{ss}} =\frac{1}{d_N}\sum_{m\in\mathcal{M}_{+}}
|J,m\rangle\langle J,m|,
\end{equation}
where $\mathcal{M}_{+}=\{-J,-J+2,\ldots,J\}$ for even $N$, whereas $\mathcal{%
M}_{+}=\{-J,-J+2,\ldots,J-1\}$ for odd $N$. Consequently, we obtain
\begin{equation}
\langle\hat J_z\rangle_{\mathrm{ss}} =%
\begin{cases}
0, & (\text{even $N$}), \\[2pt]
-\frac{1}{2}, & (\text{odd $N$}).%
\end{cases}%
\end{equation}
For even $N$, the set $\mathcal{M}_{+}$ is invariant under $m\leftrightarrow
-m$, yielding $\langle\hat{J}_z\rangle_{ss}=0$ and thus satisfying the weak
compatibility $\mathcal{I}_{xy}=0$. For odd $N$, this inversion symmetry is
absent, leaving a residual magnetization and hence $\mathcal{I}_{xy}=-1$;
the transverse parameter pair is therefore not weakly compatible.

To proceed, we consider the two-parameter unitary encoding
\begin{equation}
\hat{U}(\boldsymbol{\theta })=\exp \left[ -i\left( \theta _{y}\hat{J}%
_{y}+\theta _{z}\hat{J}_{z}\right) \right] ,\qquad \boldsymbol{\theta }%
=(\theta _{y},\theta _{z}),  \label{eq:yz_encoding}
\end{equation}%
and the parameter-dependent state $\hat{\rho}_{\boldsymbol{\theta }}=\hat{U}(%
\boldsymbol{\theta })\hat{\rho}\hat{U}^{\dagger }(\boldsymbol{\theta })$.
From Eq.~(\ref{Lalpha}), we obtain the SLD operator $\hat{L}_{y}$ for
parameter $\theta _{y}$,
\begin{equation}
\hat{L}_{y}=2i\left( \hat{P}\hat{J}_{y}\hat{Q}-\hat{Q}\hat{J}_{y}\hat{P}%
\right) ,  \label{eq:Ly_spectral_reconstruction}
\end{equation}%
which flips parity. In contrast, $\hat{L}_{z}$ preserves parity, its matrix
elements within the occupied parity sector
\begin{equation}
\langle k|\hat{L}_{z}|l\rangle =\frac{2i(\lambda _{k}-\lambda _{l})}{\lambda
_{k}+\lambda _{l}}\langle k|\hat{J}_{z}|l\rangle ,\qquad (\lambda
_{k}+\lambda _{l}>0).  \label{eq:Lz_matrix_elements}
\end{equation}

To obtain the weak compatibility condition, we calculate
\begin{equation}
\mathrm{Tr}\left( \hat{\rho}\hat{L}_{y}\hat{L}_{z}\right) =\sum_{k}\lambda
_{k}\langle k|\hat{L}_{y}\hat{L}_{z}|k\rangle =\sum_{k,l}\lambda _{k}\langle
k|\hat{L}_{y}\left\vert l\right\rangle \left\langle l\right\vert \hat{L}%
_{z}|k\rangle ,
\end{equation}%
where $\langle k|\hat{L}_{y}|l\rangle \neq 0$ requires $|l\rangle $ to have
parity opposite to that of $|k\rangle $. However, $\langle l|\hat{L}%
_{z}|k\rangle =0$, because $\hat{L}_{z}$ preserves parity. Correspondingly, $%
\langle k|\hat{L}_{y}|l\rangle \langle l|\hat{L}_{z}|k\rangle =0$, $\forall
k,l$, and thus
\begin{equation}
\mathrm{Tr}\left( \hat{\rho}\hat{L}_{y}\hat{L}_{z}\right) =\mathrm{Tr}\left(
\hat{\rho}\hat{L}_{z}\hat{L}_{y}\right) =0.  \label{smeq:LyLz}
\end{equation}%
Therefore, we obtain the weak compatibility condition between transverse and
longitudinal parameters,
\begin{equation}
\mathcal{I}_{yz}=\frac{1}{2i}\mathrm{Tr}\left( \hat{\rho}[\hat{L}_{y},\hat{L}%
_{z}]\right) =0,
\end{equation}%
and similarly $\mathcal{I}_{xz}=0$. This result is consistent with Eq.~(10)
of the main text. Moreover, the cross-sector element vanishes identically,
\begin{equation}
F_{yz}=\frac{1}{2}\mathrm{Tr}\left( \hat{\rho}\left\{ \hat{L}_{y},\hat{L}%
_{z}\right\} ]\right) =0,  \label{smeq:QFIMFyz}
\end{equation}%
in agreement with Eq.~\eqref{smeq:Fyz}, obtained from the spectral formula
for the QFIM together with the parity-symmetry selection rule. Therefore,
the two-parameter QFIM takes the diagonal form. The transverse-longitudinal
parameter pair $(\theta _{y},\theta _{z})$ is therefore automatically weakly
compatible, independently of the value of $\langle \hat{J}_{z}\rangle $ and
of the purity of $\hat{\rho}$. Unlike the two-transverse-parameter case $%
(\theta _{x},\theta _{y})$, no additional magnetization constraint is
required.

The block-diagonal form of the QFIM is a direct consequence of parity
symmetry, which eliminates all transverse--longitudinal cross terms (see Eq.~%
\eqref{smeq:LyLz}) such that $F_{xz}=F_{yz}=0$, as well as vanishing the
Uhlmann curvature $\mathcal{I}_{yz}=0$. Remarkably, this satisfies the
weak-compatibility condition, mathematically guaranteeing that both
parameter bounds can be simultaneously saturated without mutual precision
trade-offs. With projective measurements in the eigenbases of the SLD
operators, we obtain the CFIM for estimating $(\theta _{x},\theta _{y})$%
-parameter estimation [see Eqs.~\eqref{smeq:Laeigenbasis}-%
\eqref{smeq:CFIMQFIM}],
\begin{equation}
\bm{\mathcal{F}}^{(yz)}=\mathbf{F}^{(yz)}=%
\begin{pmatrix}
F_{yy} & 0 \\
0 & F_{zz}%
\end{pmatrix}%
.
\end{equation}

\subsubsection*{IV. Pure-state dynamics of the OAT Model}

The OAT model was first introduced by Kitagawa and Ueda as the mechanism to
obtain the optimal spin squeezed states~\cite{Kitagawa_supp}. Early studies
mainly focused on the reduced variance of the spin states generated from the
initial CSS $|J, \pm J\rangle_{x}$ governed by the OAT Hamiltonian $\chi
\hat{J}_{z}^{2}$ (i.e., the ideal OAT model). This initial state can be
prepared by applying a $\pi /2$ pulse to $|J,-J\rangle
=|\downarrow\rangle^{\otimes N}$. Here we consider the OAT Hamiltonian $\hat{%
H}_{\mathrm{OAT}}=\chi \hat{J}_{x}^{2}$ acting on the initial CSS $%
|J,-J\rangle$, which has been proposed to generate the GHZ state~\cite%
{MolmerPRL99_supp}.

As illustrated in Fig.~1(a) in main text, the QFI displays a prolonged
plateau characterized by $4V_{+}\approx F_{yy}\approx F_{zz}\approx N^{2}/2$%
. Beyond this plateau, the optimal phase-sensing direction switches from the
anti-squeezing direction ${\mathbf{n}}_{+}$ to the $z$ axis, as $F_{zz}$
exceeds $4V_{+}$. This behavior has been observed by numerical diagonalizing
the QFIM~\cite{Reilly_supp} and can be well understood by Eq.~%
\eqref{smeq:qfi}. Specifically, for the OAT dynamics considered here, we
obtain the pure-state solution of $F_{yy}$:
\begin{subequations}
\label{pure-state}
\begin{equation}
F_{yy}\!=\frac{N}{2}\left[ N+1-(N-1)(1-A)\right] ,
\end{equation}%
where $A=1-\cos ^{N-2}(2\chi t)$. For sufficiently large $N$, we obtain an
approximate solution of the correction:
\end{subequations}
\begin{equation}
1-A\approx e^{-(\chi t/\tau _{d})^{2}}+(-1)^{N}e^{-(\pi /2-\chi t)^{2}/\tau
_{d}^{2}},  \label{1mA}
\end{equation}%
which exhibits the ``collapse" at $\chi t=\tau _{d}$ and ``revival" at $\pi
/2-\tau _{d}$, with $\tau _{d}=1/\sqrt{2N}$. When $A=1$, the QFI reaches the
plateau value $F_{yy}=\frac{1}{2}N(N+1)$~\cite{PezzePRL09_supp}. The
approximate solution shows that the correction becomes negligibly small ($%
1-A\simeq e^{-9}$) at $\chi t=3\tau_d$ and $\chi t=\pi/2-3\tau_d$. The QFI
plateau therefore extends over the time window $3\tau_d\lesssim\chi
t\lesssim\pi/2-3\tau_d$.

Beyond the plateau, the optimal sensing direction $\mathbf{n}_{s,\max }$
swifts from ${\mathbf{n}}_{+}$ to $\hat{z}$ whenever the longitudinal QFIM
component becomes dominant, namely $F_{zz}>4V_{+}\approx F_{yy}$. Similar
result has been reported in Ref.~\cite{Reilly_supp}. To elucidate the origin
of this transition, we evaluate the difference
\begin{eqnarray}
F_{zz}-F_{yy} &=&N(N-1)(1-A)-4\langle \hat{J}_{z}\rangle ^{2}  \notag \\
&\approx &(-1)^{N}N(N-1)e^{-(\pi /2-\chi t)^{2}/\tau _{d}^{2}}-4\langle \hat{%
J}_{z}\rangle ^{2},
\end{eqnarray}%
where $1-A\approx (-1)^{N}e^{-(\pi /2-\chi t)^{2}/\tau _{d}^{2}}$ at the end
of the plateau (i.e., $\chi t\sim \pi /2-3\tau _{d}$). This expression shows
that $F_{zz}\geqslant F_{yy}$ can occur only for even $N$ case. For odd $N$,
the negative parity-dependent contribution precludes such a crossover, due
to $F_{zz}<F_{yy}$.

This parity-dependent jump in the optimal sensing direction signals the
emergence of a GHZ state in the pure-state OAT dynamics~\cite%
{MolmerPRL99_supp}. At $\chi t=\pi /2$, the OAT-evolved spin state takes the
form%
\begin{equation}
\left\vert \mathrm{GHZ}\right\rangle =\left\{
\begin{array}{cc}
\!\frac{e^{-i\pi /4}}{\sqrt{2}}\left( \left\vert J,-J\right\rangle +e^{i\pi
(N+1)/2}\left\vert J,J\right\rangle \right) , & (\text{even $N$}) \\[8pt]
\!\frac{e^{i7\pi /8}}{\sqrt{2}}\left( \left\vert J,-J\right\rangle
_{y}+e^{i\pi N/2}\left\vert J,J\right\rangle _{y}\right) , & (\text{odd $N$})%
\end{array}%
\right.
\end{equation}
where $|J,m\rangle _{y}=e^{-i\frac{\pi }{2}\hat{J}_{x}}|J,-m\rangle $ denote
eigenstates of $\hat{J}_{y}$. These highly entangled states exhibit a
parity-dependent optimal sensing direction. For a single-parameter
phase-encoding with even $N$, the phase sensing along the $z$ axis is
optimal as $F_{zz}=N^{2}$ ($F_{yy}=N$), whereas for odd $N$, the sensing
along $y$ axis is optimal as $F_{yy}=N^{2}$ ($F_{zz}=N$).

\section*{References for Supplemental Material}

\end{document}